\documentclass[onecolumn]{IEEEtran}%

\usepackage[utf8]{inputenc}
\usepackage{amssymb}
\usepackage{lipsum}
\usepackage[intlimits]{amsmath}
\usepackage{amsfonts}
\usepackage{amsthm}
\usepackage{mathptmx}
\usepackage{graphicx}
\usepackage{makeidx}
\usepackage{fancyhdr}
\usepackage{lettrine}
\usepackage{mathpazo}

\usepackage{avant}
\usepackage{microtype}
\usepackage{color}%T
\usepackage{setspace}
\usepackage{enumitem}

\usepackage{listings}
\usepackage{algorithm}
\usepackage{algorithmicx}
\usepackage{algpseudocode}

\algnewcommand\algorithmicoutput{\textbf{Output:}} 
\algnewcommand\Output{\item[\algorithmicoutput]}
\algnewcommand\algorithmicinput{\textbf{Input:}} 
\algnewcommand\Input{\item[\algorithmicinput]}

\usepackage[table]{xcolor}

\usepackage{framed}
\definecolor{shadecolor}{gray}{0.9}

\newcounter{theExample}
\newcounter{theRemark}
\newenvironment{Remark}{
	\par\smallskip\refstepcounter{theRemark}%
	\noindent%
	\textbf{Remark~\arabic{theRemark}}:~%
	\ignorespaces%
}{
	\par\smallskip
}

\newtheorem{exmplA}{Example}
\newenvironment{exmpl}
{\begin{shaded}\begin{exmplA}\small\hangindent=7mm \addtolength{\parindent}{7mm}}
		{\end{exmplA}\ignorespacesafterend\end{shaded}\ignorespacesafterend}
	
\begin{document}

	\title{Convolutional Neural Networks Demystified: A Matched Filtering Perspective Based Tutorial}

	\author{
		Ljubi\v{s}a~Stankovi\'{c},~\IEEEmembership{Fellow,~IEEE,}
		Danilo Mandic,~\IEEEmembership{Fellow,~IEEE,}
		\thanks{
			L. Stankovi\'{c} is with the  University of
			Montenegro, Podgorica, Montenegro. D. Mandic is with the Imperial College London, London, United Kingdom.
			Contact e-mail: ljubisa@ucg.ac.me

\bigskip

}

	}
	\maketitle

	\setcounter{tocdepth}{3}

	\begin{abstract}
Deep Neural Networks (DNN) and especially Convolutional Neural Networks (CNN) are a de-facto standard for the analysis of large volumes of signals and images. Yet, their development and underlying principles have been largely performed in an ad-hoc and black box fashion. To help demystify CNNs, we revisit their operation from first principles and a matched filtering perspective. We establish that the convolution operation within CNNs, their very backbone, represents a matched filter which examines the input signal/image for the presence of pre-defined features. This perspective is shown to be physically meaningful, and serves as a basis for a step-by-step tutorial on the operation of CNNs, including pooling, zero padding, various ways of dimensionality reduction. Starting from first principles, both the feed-forward pass and the learning stage (via back-propagation) are illuminated in detail, both through a worked-out numerical example and the corresponding visualizations. It is our hope that this tutorial will help shed new light and physical intuition into the understanding and further development of deep neural networks.
	\end{abstract}

\bigskip

{ \large \bf This work has been submitted to the IEEE for possible publication. Copyright may be transferred without notice, after which this version may no longer be accessible.}

\bigskip

\section{Convolutional Neural Network -- CNN}

We live in a world overwhelmed by data with multiple sources routinely generating high resolution signal/image streams. Processing such data comes with an inevitable bottleneck of the (curse of) dimensionality. To put this into perspective, even a modest resolution $640 \times 480$ VGA image comprises 307,200 pixels, while a $1920 \times 1080$ HDTV image contains 2,073,600 pixels; if those images are processed using neural networks (NN), then each pixel requires one neuron at the NN input layer. This is followed by at least one hidden layer, so that even for a typical small-scale fully connected hidden layer with, for example, 1024 nodes, the number of parameters quickly becomes prohibitive \cite{hassoun1995fundamentals,yegnanarayana2009artificial,gurney2018introduction}. 

In practical applications, this issue is partially mitigated by exploiting the fact that most physical data  sources exhibit a smooth nature, so that the neighboring signal points or image pixels exhibit some sort of similarity. This allows us to employ local information in the form of  features, which describe the analyzed signals/images; our task then becomes to search for specific localized features in data, instead of the standard brute force approaches which scale exponentially with the data volume. 

Another advantage of operating in the feature space, instead of with the raw pixels, is that this resolves the important problem of position change of the patterns in data due, for example, to translation. Namely, if a certain data feature changes its position, then a pixel-wise approach will assume a complete change in pixels, while a feature-wise approach will look for specific shapes anywhere in the signal. 

Similar reasoning also underpins  the operation of convolutional neural networks (CNNs), which boils down to performing some sort of search for features in the analyzed signal, such that these features are invariant to their position change \cite{kuo2016understanding,mandic2001recurrent,kiranyaz20211d,kuo2017cnn,ghosh2020fundamental,li2016survey,zhang2019recent,albawi2017understanding,o2015introduction,jin2017deep,dong2016accelerating,acharya2017deep,kim2017convolutional}. 
More specifically, the window used in convolution within CNNs (\textit{convolution filter or convolution kernel}) is capable of recognizing precisely one feature that is matched to its form. To do so, we perform feature matching over the whole signal, akin to a mathematical lens in search of specific forms. For more details on a similar approach to graph convolutional neural networks, please se our sister paper \cite{GCNN_tutorial}. 

\section{Principle of Matched Filtering} While the use of a convolutional window has become a de facto standard in CNNs, an open question remains of how we can justify that the convolution is an appropriate operation for detecting features in a signal -- a subject of this tutorial. To this end, we draw inspiration from the matched filter theory, whereby the convolution of the considered signal with the feature that we are looking will confirm the existence and location of the feature at hand. Recall that the output of a matched filter is calculated through a convolution \cite{stankovic2015digital}
$$y(n)=x(n)*w(-n)=\sum_{m}x(m)w(m-n)=\sum_mx(n+m)w(m)=x(n)*_cw(n),$$ where $w(n)$ denotes the feature that we are looking for in the analyzed signal $x(n)$ and $*_c$ denotes the convolution with the time-reversed feature/template,  $w(-n)$, which serves as the ``impulse response''. Therefore, the best search function to detect a feature, $w(n)$, within a signal $x(n)$, would be through a convolution of the signal $x(n)$ with $w(-n)$. 

\begin{Remark}
Convolution based feature detection produces a result that is independent of the feature position within the considered signal, since 
$y(n)=x(n)*w(-n)$
is 
calculated by sliding the window (filter/kernel), $w(n)$, and multiplying it with the signal segments, $x(n)$ for all $n$. The same holds for an image, whereby when calculating the corresponding convolution with $w(-m,-n)$ a two-dimensional filter $w(m,n)$ slides along the image in both spatial directions. 

	If we are looking for one of $K$ features in the input signal then we can form a bank of $K$ matched filters with outputs
	$$y_k(n)=x(n)*w_k(-n)=\sum_mx(n+m)w_k(m)=x(n)*_cw_k(n), \ \ k=1,2,\dots,K.$$
	
	The decision about the feature $k$ which is contained in the input signal is based on 
	$$k=\textrm{arg}\{\max \{x(n)*_cw_1(n), \ x(n)*_cw_2(n), \dots, x(n)*_cw_K(n) \}\}.$$

\end{Remark}

\begin{figure}
\begin{center}
	\includegraphics[scale=0.9]{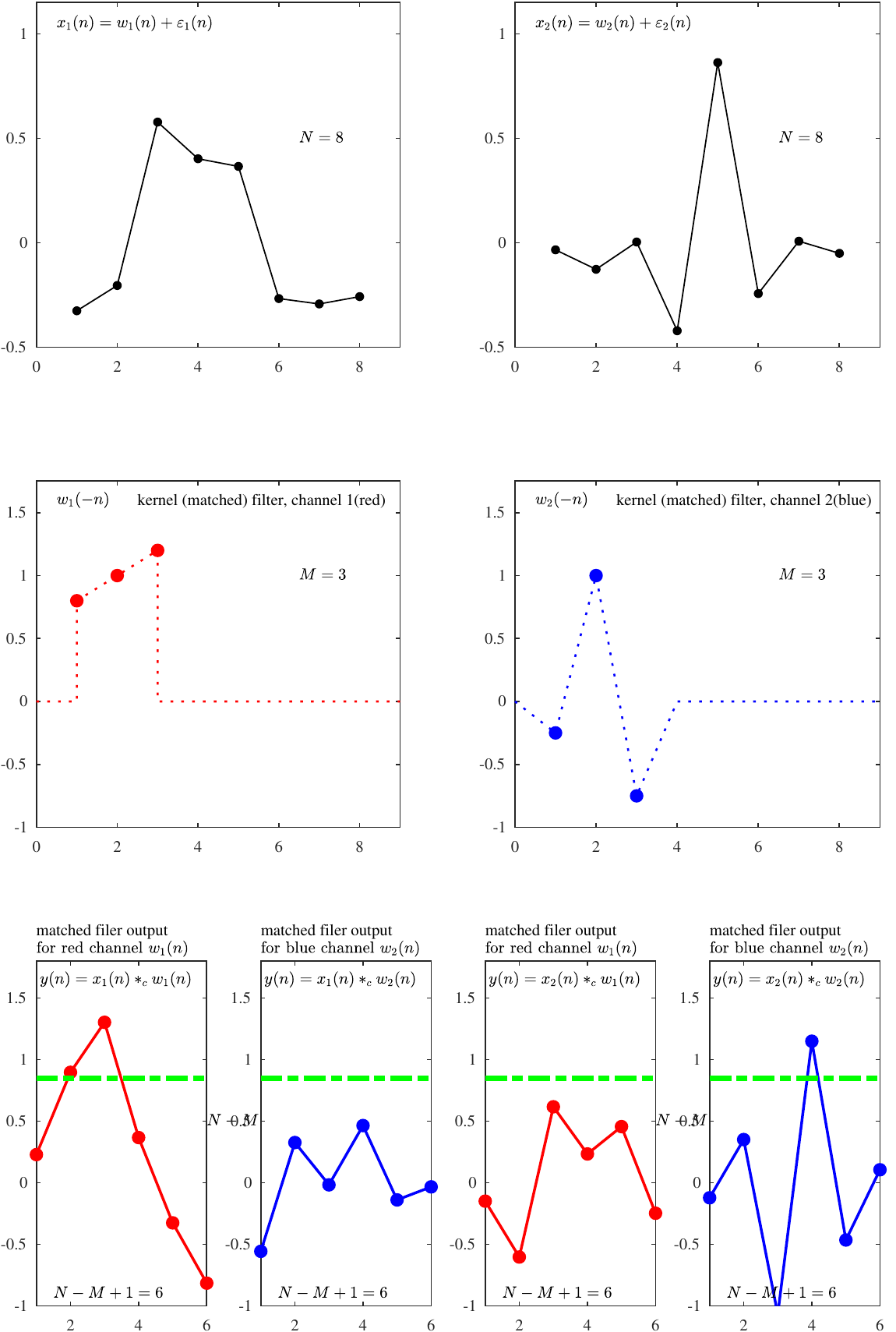} 
	\caption{Illustration of the matched filter operation. Top panels: Two noisy input signals of the length $N=8$ (noise is denoted by $\varepsilon(n)$). Middle panels: Two features that we are searching for in the input signals of the length $M=3$. The reversed versions of these features serve as the inputs to the corresponding (red and blue channels) matched filters (middle panels). Bottom panels: Deep Neural Networks (DNN) and especially Convolutional Neural Networks (CNN) are a de-facto standard for the analysis of large volumes of signals and images. Yet, their development and underlying principles have been largely performed in an ad-hoc and black box fashion. To help demystify CNNs, we revisit their operation from first principles and a matched filtering perspective. We establish that the convolution operation within CNNs, their very backbone, represents a matched filter which examines the input signal/image for the presence of pre-defined features. This perspective is shown to be physically meaningful, and serves as a basis for a step-by-step tutorial on the operation of CNNs, including pooling, zero padding, various ways of dimensionality reduction. Starting from first principles, both the feedforward pass and the learning stage (via backpropagation) are illuminated in detail, both through a worked-out numerical example and the corresponding visualisations. It is our hope that this tutorial will help shen new light and physical intuition into the understanding and further development of deep neural networks.The outputs of the read and blue matched filter to the each of input signals, with appropriate  decision line, used to decide which feature is contained in the corresponding input signal. }%
	\label{Example_CNN_Matched}%
\end{center}
\end{figure}

\begin{exmpl}
The principle of the matched filter is illustrated on two noisy signals shown in the two top panels in Fig. \ref{Example_CNN_Matched}. The features contained in these signals are given in the two middle panels in Fig. \ref{Example_CNN_Matched} and are designated by the red and blue lines. Both signals are convolved with the reversed versions of these two features (serving as the matched filter impulse responses) according to $y(n)=x(n)*w(-n)$. The outputs of the matched filters (red and blue filter) are shown in the bottom panels in Fig. \ref{Example_CNN_Matched}. The left two panels at the bottom show the output of the red and blue matched filter to the first input signal at the top-left, while  the right two panels at the bottom show the output of the red and blue matched filter to the second input signal at the top-right. We can conclude that the first input signal contains the red feature (since the output is above the threshold line), while the second input signal contains the blue feature. 
\end{exmpl}

\begin{figure}
	\centering
	
	\includegraphics[scale=0.9]{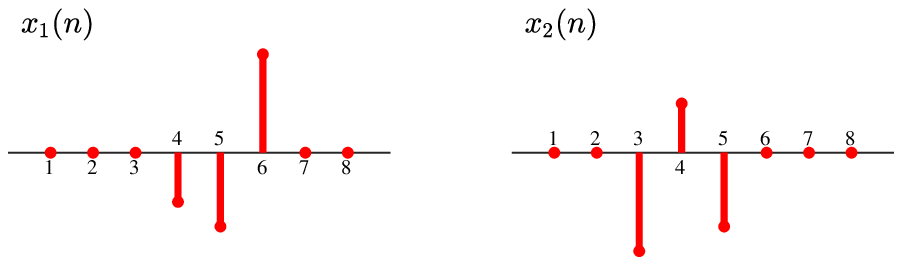} \\
	(a) \hspace{65mm} (b)
	\\ \vspace{3mm} 
	\includegraphics[scale=0.9]{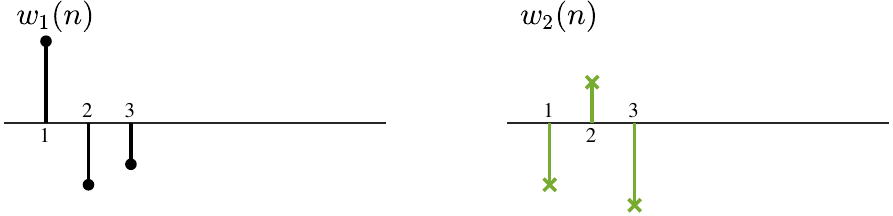} \\
	(c) \hspace{65mm} (d)
	\\ \vspace{7mm} 
	\includegraphics[scale=0.9]{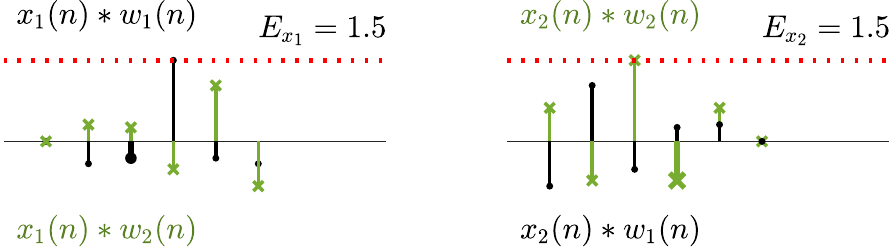}	 \\
	(e) \hspace{65mm} (f)
	\caption{Example of matched filtering, with two input signals, $\mathbf{x}_1$ and $\mathbf{x}_2$, (containing different features) observed in two different scenarios.  (a) The input signal with the first feature, ${\bf x}_{1}$. (b) The second input signal. 
		The impulse responses of the corresponding matched filters are shown respectively in panels (c) and (d).	
		The outputs of the graph matched filters, with the  impulse response $w_1(n)$ corresponding to the feature $x_1(n)$  and $w_2(n)$ corresponding to $x_2(n)$, are given respectively in panels (e) and (f). Observe from panel (e) that, as desired,  matched filter ${\bf w}_{1}$ correctly detected the presence of feature ${\bf x}_{1}$ (black line), with the maximum output at $n=5$, while the matched filter ${\bf w}_{2}$ failed to detect feature ${\bf x}_{1}$ as it was not designed for this purpose (green line). Panel (f) depicts an analogous scenario for the detection of feature ${\bf x}_{2}$, with the matched filter ${\bf w}_{2}$ performing a correct detection with the maximum output at $n=3$. The maximum values of the outputs of the two  matched filters (energies of the corresponding input features) are marked by larger symbols and a dotted read line, while their respective values  are denoted by $E_{x_1}$ and $E_{x_2}$.}
	\label{fig-sig-gr1}
\end{figure}

Notice that in the definition of the matched filter, the convolution $x(n)*w(-n)$ corresponds to the cross-correlation of $x(n)$ and $w(n)$ rather than their convolution, $x(n)*w(n)$. This because we have used a digital filter to implement convolution, which has been made possible by the impulse response being a time-reversed version of the feature. Nonetheless, the network is called the \textit{convolutional neural network} (CNN) rather than the \textit{cross-correlational neural network}, with all notations assuming that the convolution is applied after one of the signals is time reversed, that is $x(n)*w(-n)$. This is implicitly indicated in various notations in literature, for example,  $\mathbf{x}*rot 180^0 (\mathbf{w})$ or $\textrm{conv}(\mathbf{x}),reverse(\mathbf{w})$. We will use a simplified notation $\mathbf{x}*_c\mathbf{w}$, to indicate that the second signal in the convolution is reversed.

\begin{Remark} Consider receiving a waveform which is one from a set of possible waveforms (dictionary) for our problem. The task is to determine which of the template waveforms it matches best. Then, it intuitively makes sense to compute the correlation of the received waveform against each member of the alphabet. Assuming the same normalized energy, the maximum correlation occurs against the correct template waveform from the dictionary. One way of calculating this correlation is by putting the received waveform through a bank of filters  with each having as an impulse response one of the alphabet signals, time reversed. Then, the  maximum output value will be equal to the cross-correlation of the received signal with the alphabet signal. 
\end{Remark}
	
	 CNNs are a type of neural network that use convolution layers, which consist of a set of convolutional filters. Convolutional filters are typically applied over different layers, each aiming to identify a different feature in a signal.  By learning different forms of the feature space, convolutional networks allow for robust, efficient analysis and classification of signals and images.

\section{The Forward Propagation Pathway in CNNs}\label{forwardP}

We shall now use a matched filter perspective to shed a new light on key algorithmic steps in the operation of CNNs, for simplicity we assumed  already (initialized) or calculated weights of the convolutional filters (\textit{forward propagation}). The weight update will be addressed afterwards.

\begin{enumerate}
	
	\item \textbf{Input:}  Consider a signal, $\mathbf{x}$, with $N$ samples, given by
	$$\mathbf{x}=[x(0), \ x(1), \ \dots, x(N-1)]^T$$
	(or an image of $N\times N$ samples) as the input to a neural network. 
	
	A common goal in CNNs is to classify input signals (images) into several non-overlapping sets (clusters).
	
	\bigskip
	
	\item \textbf{Convolution layer:} This operation employs a convolutional filter of $M$ elements (\textit{cf.} a filter of $M\times M$ samples for images). The convolution layer filter is sometimes called convolutional kernel. Common choices are, for example, those of length $M=3$ or $M=5$. Note that $K$ such filters are applied, if we are looking for $K$ features in $\mathbf{x}$. The  elements of the $k$-th response of the first convolutional layer are then 
	$$\mathbf{w}_k^1=[w_k^1(0), \ w_k^1(1), \ \dotsc, \ w_k^1(M-1)]^T,$$
	for $k=1,2,\dots,K$.
	 Then, the output signals are
	$$\mathbf{y}_k^1=\mathbf{x}*_c\mathbf{w}_k^1,$$
	where $*_c$ denotes the convolution  of the time-reversed filter (channel), $\mathbf{w}_k^1$, and the  signal, $\mathbf{x}$, (cross-correlation). For further illustration, the element-wise form of this convolution, for $M=3$, is given by
	\begin{equation}y_k^1(n)=w_k^1(0)x(n)+w_k^1(1)x(n+1)+w_k^1(2)x(n+2)=\sum_{m=0}^{M-1}w_k^1(m)x(n+m).
	\label{ConvCNN3}
	\end{equation}
	The dimension of the $k$th output, $\mathbf{y}_k^1$, is $(N-M+1)\times 1$. The last element in $y_k^1(n)$ is obtained for $(n+m)=N-1$ with $m=M-1$, that is,  $y_k^1(N-M)$. For $M=3$, the last element in $y_k^1(n)$ is $y_k^1(N-3)$. 
	
	In total, $K$ such output signals of the convolution layer, $\mathbf{y}_k^1$, $k=1,2\dots,K$, are obtained, with the total number of the output signal elements, $y_k^1(n)$, from the first convolution layer therefore being $K(N-M+1)$.  
	
	\begin{Remark}
	The total number of filter weights, $w_k^1(n)$, in the first convolutional layer is equal to the product of the convolution filter length, $M$, and the number of filters, $K$, that is $MK$. This is typically much smaller than in the case of a fully connected neural network, whereby each of $N$ input signal samples is connected through weights to each of $K$ output signals, to yield $NK$ connections. 
	\end{Remark}
	
	If an image is considered, then the output image of the convolutional filter is of the size $(N-M+1)\times(N-M+1)$. There are $K$ such images and the total number of the filter weights is $KM^2$, which is again smaller than $KN^2$ connections in the standard fully connected layer.
	
	\begin{exmpl}
		\textbf{Relation to standard neural networks.} The input-output relation for the CNN simplifies into standard neural network as a special case. This is immediately seen by first considering the element-wise form of the output at an unindexed neuron (before the activation function), given by
		\begin{equation}
		y(n)= \sum
		_{k=1}^{N}w_{k}x_{k}(n) \label{ConvNNNN}
		\end{equation}
		where $x_{k}(n)$ designates the input to the neuron $k$ in a time instant, $n$. 
		
		In the CNN, one signal, $x(n)$,  is considered as the input whose $N$ samples arrive simultaneously at $N$ input neurons (these are considered as one training  datum). Such data are  connected to $K$ output neurons in the first convolutional layer, so that the input-output relation within the CNN framework becomes
		\begin{equation}
		y_k=  \sum
		_{m=0}^{N-1}w^1_k(m)x(m)=w^1_{k}(0)x(0)+w^1_{k}(1)x(1)+\dots+w^1_k(N-1)x(N-1). \label{ConvNNN}
		\end{equation}
	
		Next, by comparing (\ref{ConvNNNN}) and its CNN notation in (\ref{ConvNNN}) to (\ref{ConvCNN3}) we can conclude that the standard neural network is a \textbf{special case} of the CNN, with $M=N$. Namely, the relation in (\ref{ConvCNN3}), for $M=N$ is of the form
		\begin{equation}y_k^1(n)=w_k^1(0)x(n)+w_k^1(1)x(n+1)+w_k^1(2)x(n+2)+\dots+w_k^1(N-1)x(n+N-1). \nonumber
		\end{equation}
		If zero-padding is not performed for the input signal, this relation can be calculated only for $n=0$ (since $x(n)$ is defined only for $n=0,1,\dots, N-1$), to yield 
		\begin{equation}y_k^1=y_k^1(0)=w_k^1(0)x(0)+w_k^1(1)x(1)+w_k^1(2)x(2)+\dots+w_k^1(N-1)x(N-1), \text{ for } k=1,2,\dots,K,
		\label{ConvCNN3N1}
		\end{equation}
		 so that we arrived at (\ref{ConvNNN}). For this reason, $M \ll N$ is typically used in CNNs. See Fig. \ref{fully_connected_layer1} for a step-by-step illustration of the operation of the first convolutional layer. 
		
		\medskip
		
		\noindent\textit{Note:} All results derived next for the convolutional layer also hold for the standard, fully connected layer, with $y_k^1=y_k^1(n)$, and using only $n=0$, and $M=N$.
		
		In the standard neural network, when the input $x(n)$, $n=0,1,\dots,N-1$, is simultaneously applied to $N$ input neurons that are connected to $K$ output neurons, with $y_k$, $k=1,2,\dots,K$, the number of different weights, $w^1_k(n)$, is  $NK$, which is larger than $MK$, the number of weights in the CNN.
	
\end{exmpl}
\begin{figure}[hptb]
\begin{center}
	\includegraphics{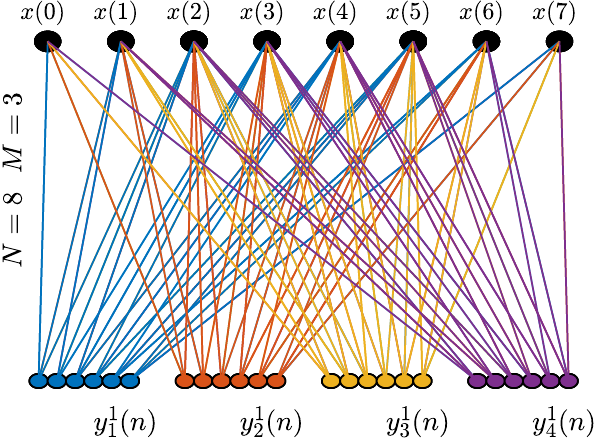} \hspace{3mm} \includegraphics{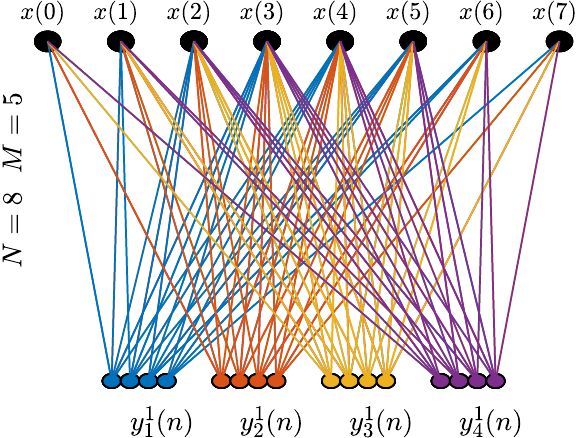} \\ \vspace{5mm} \includegraphics{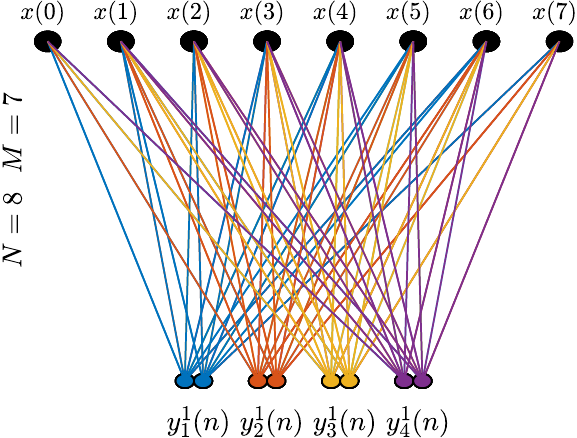} \hspace{3mm} \includegraphics{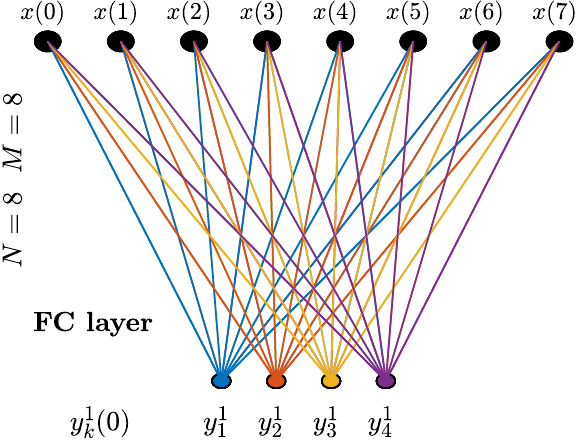}
	\caption{Operation  of a Convolutional Neural Network for an input of size $N$, convolution layer with $K=4$ output neurons, and for various lengths, $M$, of the convolution filter. For $M=N$, the convolutional layer reduces to the fully connected (FC) layer. Top left: The case for $N=8$ and $M=3$.  Top right: The case for $N=8$ and $M=5$.  Bottom left: The case for $N=8$ and $M=7$.  Bottom right: The case for $N=8$ and $M=d8$, corresponding to the FC layer.}%
	\label{fully_connected_layer1}%
\end{center}
\end{figure}    
	
\bigskip

	\item \textbf{Bias:}  In the convolution layer (like in  standard neural network layers), the bias (constant) term may be added, to yield 
	$$y_k^1(n)=\sum_{m=0}^{M-1}w_k^1(m)x(n+m)+b_k^1.$$
	The vector form of the CNN output then becomes
	$$\mathbf{y}_k^1=\mathbf{x}*_c\mathbf{w}_k^1+b_k^1,$$
	with the total number of coefficients into every convolution increased by one. 
	
	\begin{Remark}
	\textit{The number of weights (parameters) in a CNN  depends only on the size of the convolutional (feature matching) filter.} For time-domain signals, with  the bias term included, the total number of weighs is  $K(M+1)$.  For an image, the total number of weights is $K(M^2+1)$.
	\end{Remark}
	
\bigskip
	
	\item \textbf{Zero-Padding:} In some cases, the output of the convolution stage should be of the \textit{same} size as the input signal (image), instead of the minimum of $(N-M+1)$ which results from the convolution. This can be achieved if the input signal (image) is zero-padded with an appropriate number of zeros.  For example, for $M=3$ and any $N$, we could add a zero at $x(-1)=0$ and a zero at $x(N)=0$, to calculate the convolution output $y(n)$ as 
	$$y_k^1(n)=w_k^1(0)x(n-1)+w_k^1(1)x(n)+w_k^1(2)x(n+1).$$
	which yields the same number  of samples as in the input signal, $x(n)$. Namely, now we can calculate $y(0)$ as the first element in the convolution, and $y(N-1)$ as the last element in the convolution. 
	
	\begin{Remark}
		In general, if the filter response length is $M$, the signal should be padded with $(M-1)$ zeros to perform a convolution of length $N$, as the length of the convolution operation is $N-M+1$, if the filter response should not be moved outside the signal. For a filter with an odd number of elements $M$, if we desire that the convolution result has the same number of samples as the input signal, the input signal could symmetrically be zero-padded, by $(M-1)/2$ elements before the starting sample at $n=0$, and with the same number of zero elements after the signal sample at $n=N-1$.
		\end{Remark}
	
\bigskip
	
	\item \textbf{Nonlinear activation function:} Signals and images are far from exhibiting a linear nature, while convolution (correlation) is a linear operation. To this end, a non-linearity is applied to the output of a convolutional layer. Such a nonlinear map will restrict the output values to reside within a specified output range, as in the case of sigmoid type of nonlinearities. The most common  nonlinear activation function for CNNs is the Rectified Linear Unit (ReLU), defined by 
	$$f(x) = \max\{0,x\}.$$
	In the CNNs, this function has several advantages over sigmoidal-type activation functions: (i) It does not saturate for the positive values of its input thus producing nonzero gradient for large input values, (ii) Its calculation is not computationally demanding, and (iii) In practical applications ReLU converges faster than the saturation-type nonlinearities (logistic, tanh).  Moreover, this function does not activate all neurons at the same time, so that sparsification by deactivation is achieved for each neuron producing negative value as an input to the ReLU activation function.       
	
	The output of one convolutional layer, after the activation function, then becomes
	$$f(\mathbf{y}_k^1)=f(\mathbf{x}*_cmathbf{w}_k^1+{b}_k^1)=f(\mathbf{w}^1_k,\mathbf{x}).$$
	For our example with $M=3$, the element-wise output is therefore
	$$f(y_k^1(n))=f\Big(w_k^1(0)x(n)+w_k^1(1)x(n+1)+w_k^1(2)x(n+2)+b_k^1\Big).$$ 
	
	\begin{exmpl}\label{reLUex}
		Consider a multivariate signal $\mathbf{y}_k=\mathbf{x}*_c\mathbf{w}_k^1+b_k$, with $K=3$ convolution filters (channels), $k=1,2,3$ given by  
		
		\noindent$\mathbf{y}_1=\begin{bmatrix}
		\phantom{-}0.35 & \phantom{-}0.49 & -0.65 & -0.65 & -0.69 & \phantom{-}0.48 \end{bmatrix}^T
		$ \\
		$\mathbf{y}_2=\begin{bmatrix}
		-0.05 & -0.06 & -0.28 & -0.21 & \phantom{-}0.13 & \phantom{-}0.37 \end{bmatrix}^T
		$  \\
		$\mathbf{y}_3=\begin{bmatrix}
		\phantom{-}0.48 & \phantom{-}0.50 & -0.77 & -1.66 & -0.76 & \phantom{-}0.71 \end{bmatrix}^T.$
		
		The element-wise output from the ReLU activation function is then 
		
		\noindent$f(\mathbf{y}_1)=\begin{bmatrix}
		0.35 & 0.49 & \textbf{0.00} & \textbf{0.00} & \textbf{0.00} & 0.48 \end{bmatrix}^T$ 
		\\
		$f(\mathbf{y}_2)=\begin{bmatrix}
		\textbf{0.00} & \textbf{0.00} & \textbf{0.00} & \textbf{0.00} & 0.13 & 0.37 \end{bmatrix}^T$  \\
		$f(\mathbf{y}_3)=\begin{bmatrix}
		0.48 & 0.50 & \textbf{0.00} & \textbf{0.00} & \textbf{0.00} & 0.71 \end{bmatrix}^T,$ 
		
		It is also convenient to consider the indicator matrix  that designates the active/deactivated neurons after the ReLU activation, which is  given by 
		$$\mathbf{M}^{ReLU}=\arraycolsep3pt\begin{bmatrix}
		1  & 1 &0  &0  &0  &1  \\
		0 &0  &0  &0  & 1  &1 \\
		1  &1 &0 &0  &0  & 1
		\end{bmatrix}^T.$$
		
	\end{exmpl}         
	
	Since the ReLU is defined in such a way that it produces a zero output for negative input values, the main problem with the ReLU activation function arises when the input to a  neuron has many negative values so that the corresponding zero-output of the ReLU function will leave many neurons without update of their weights (“dying ReLU”). This problem can be avoided using Leaky ReLU, whereby negative values of the input are mapped onto small scaling factors, for example, $f(y_k(n))=0.01y_k(n)$, for $y_k(n)<0$.
	
\bigskip
	
	\item \textbf{Stride -- Convolution step (down-sampling):} In calculating the convolution, it is common that the convolution filter is shifted along by one step, so that  after $y_k(n)$ is obtained at an instant $n$, the next convolution is calculated at $(n+1)$. However, if the signal is sufficiently dense and slow-varying, for computational reasons we may decide to skip several time instants before the next convolution, $y_k(n)$, is calculated. This operation effectively represents downsampling  of the output signal in the convolution calculation, whereby the degree of downsampling is called the \textit{stride} (step). For example, if  one time instant (or pixel in both directions) is skipped before the next calculation of the convolution value, then the stride value is equal to two, which corresponds to the down-sampling the convolution output, $y_k(n)$,  by a factor of 2. The stride value of four, which would mean that the convolution $y_k(n)$ is calculated at every fourth instant (pixel) of the original signal $x(n)$.
	
	\begin{exmpl}
		Consider the output from the ReLU activation function in  Example \ref{reLUex}, given by 
		
		\noindent$f(\mathbf{y}_1)=\begin{bmatrix}
		\textbf{0.35} & 0.49 & 0.00 &\textbf{ 0.00} & 0.00 & 0.48 \end{bmatrix}^T$ 
		\\
		$f(\mathbf{y}_2)=\begin{bmatrix}
		\textbf{0.00} & 0.00 & 0.00 & \textbf{0.00} & 0.13 & 0.37 \end{bmatrix}^T$  \\
		$f(\mathbf{y}_3)=\begin{bmatrix}
		\textbf{0.48} & 0.50 & 0.00 & \textbf{0.00} & 0.00 & 0.71 \end{bmatrix}^T.$ 
		
		The output with stride 3 is obtained by down-sampling the outputs $f(\mathbf{y}_k)$ by the factor of 3 to give 
		\noindent$\mathrm{Stride}_3\{f(\mathbf{y}_1)\}=\begin{bmatrix}
		0.35 &  0.00 \end{bmatrix}^T$ 
		\\
		$\mathrm{Stride}_3\{f(\mathbf{y}_2)\}=\begin{bmatrix}
		0.00 & 0.00  \end{bmatrix}^T$  \\
		$\mathrm{Stride}_3\{f(\mathbf{y}_3)\}=\begin{bmatrix}
		0.48  & 0.00  \end{bmatrix}^T.$ 
		
		The indicator matrix which corresponds to upsampling (inserting zeros) from $\mathrm{Stride}_3\{f(\mathbf{y}_k)\}$ to the original solution  $f(\mathbf{y}_k)$ is then given by 
		$$\mathbf{M}^{Stride_3}=\arraycolsep3pt\begin{bmatrix}
		1  & 0 &0  &1  &0  &0  \\
		1 &0  &0  &1  & 0  &0 \\
		1  &0 &0 &1  &0  & 0
		\end{bmatrix}^T.$$
		
	\end{exmpl}

	\bigskip
	
	\item \textbf{Pooling:} In order to reduce the possibly excessive size of the data throughput, the output signals at each layer are typically further down-sampled through the so called \textit{pooling} operation, in addition to the stride type of down-sampling scheme described above. A typical pooling operation of $y_k(n)$ in a CNN is the max-pooling which splits the output signal into nonoverlapping segments of  $P$ samples and takes the maximum value from each segment (for an image, we split the image into $P \times P$ nonoverlapping segments and take the maximum value from every such segment).  The signal at the output of the \textit{max-pooling} step with $P$ segments then becomes
	$$o_k^1(n)=\max\{f(y_k(n)),f(y_k(n+1)),\dots,f(y_k(n+P-1))\}=F_1\Big(x(n),w(n)\Big)$$
	or in a vector form
	$$\mathbf{o}_k^1=F_1(\mathbf{w}_k^1,\mathbf{x}).$$
	
	The max-pooling reduces the  size of the representation, and thus helps decrease of the  computation requirements and the number of weights in a CNN. Pooling also provides some translation invariance, since it choses the maximum value among $P$ neighboring samples, regardless of their position. Other forms of pooling include the average-pooling, whereby the output is an average of $P$ neighboring samples.
	
	\begin{exmpl}
		Consider the output from the ReLU activation function in  Example \ref{reLUex}.
		Then, the output from the max-pooling operation, with $P=3$, is obtained from $f(\mathbf{y}_k)$ as	
		$$\mathbf{o}^1=\begin{bmatrix} \max\{	0.35 & 0.49 & 0.00\} & \max\{ 0.00 & 0.00 & 0.48 \} \\
		\max\{ 0.00 & 0.00 & 0.00 \} & \max\{ 0.00 & 0.13 & 0.37 \} \\
		\max\{ 0.48 & 0.50 & 0.00 \} & \max\{ 0.00 & 0.00 & 0.71\}\end{bmatrix}^T=\begin{bmatrix}
		0.49 &0.48 \\
		0.00 &0.37 \\
		0.50 &0.71 
		\end{bmatrix}^T$$ 
		
		The corresponding indicator matrix for the upsampling from the downsampled $\mathbf{o}^1$ to the original size of $f(\mathbf{y}_k)$ is given by 
		$$\mathbf{M}^{MP}=\arraycolsep3pt\begin{bmatrix}
		0  & 1 &0  &0  &0  &1  \\
		1 &0  &0  &0  & 0  &1 \\
		0  &1 &0 &0  &0  & 1
		\end{bmatrix}^T.$$
		
		Notice that the max-pooling with $P=3$ reduces the size of $f(\mathbf{y}_k)$ from $6$ to $2$, the same as when employing the stride factor of 3. However, unlike  in the stride operation case, in the max-pooling the positions of the selected samples, and the corresponding upsampling matrix, are signal dependent.
		
	\end{exmpl}

	\bigskip

	\item \textbf{Flattening:} The one-dimensional signals $\mathbf{o}^1$, after pooling, are already in a vector form. These vectors are then concatenated to form the vector with the elements
	$$o^1_F\Big((k-1)(N-M+1)+n\Big)=o^1_k(n), \text{    for } k=1,2,\dots,K \text{ and } n=0,1,\dots,N-M. $$
	This  vector is of size $K(M-M+1)$ if no max-pooling is performed. If max-pooling with a factor of $P$ is used, the size of the concatenated (\textit{flattened}) vector $\mathbf{o}^1_F$ is $K(M-M+1)/P$ . 
	
	In the case of images, while after the max-pooling the output still remains a two-dimensional object; these images are also rearranged into the vector form and concatenated into one vector. 
	
	This process is called the \textit{flattening} operation. 
	
	\begin{exmpl}\label{cnn_ex1}
		Fig. \ref{CNN_conv} depicts the operation of the first convolution layer of a CNN for an input signal of $N=32$ samples each, four convolution filters ($K=4$) with $M=5$ samples, 
		 the ReLU nonlinear activation function $\max\{0,y^1_k+b^1_k\}$, $k=1,2,3,4$, and max pooling with factor $P=2$. At the max-pooling stage,  the signal is grouped into segments of two samples and the largest sample  then represents the output of this operation. The output from the max-pooling stage is then either used as input to the next convolution layer (in the case of multiple convolutional layers) or it is flattened  and fed to the neurons of a common fully connected (FC) neural network.
		The initial weights of the convolution filter are generated as Gaussian random numbers (common way for CNN initialization).
	\end{exmpl}
	
	\begin{figure}[hptb]
		\begin{center}
			\includegraphics[]{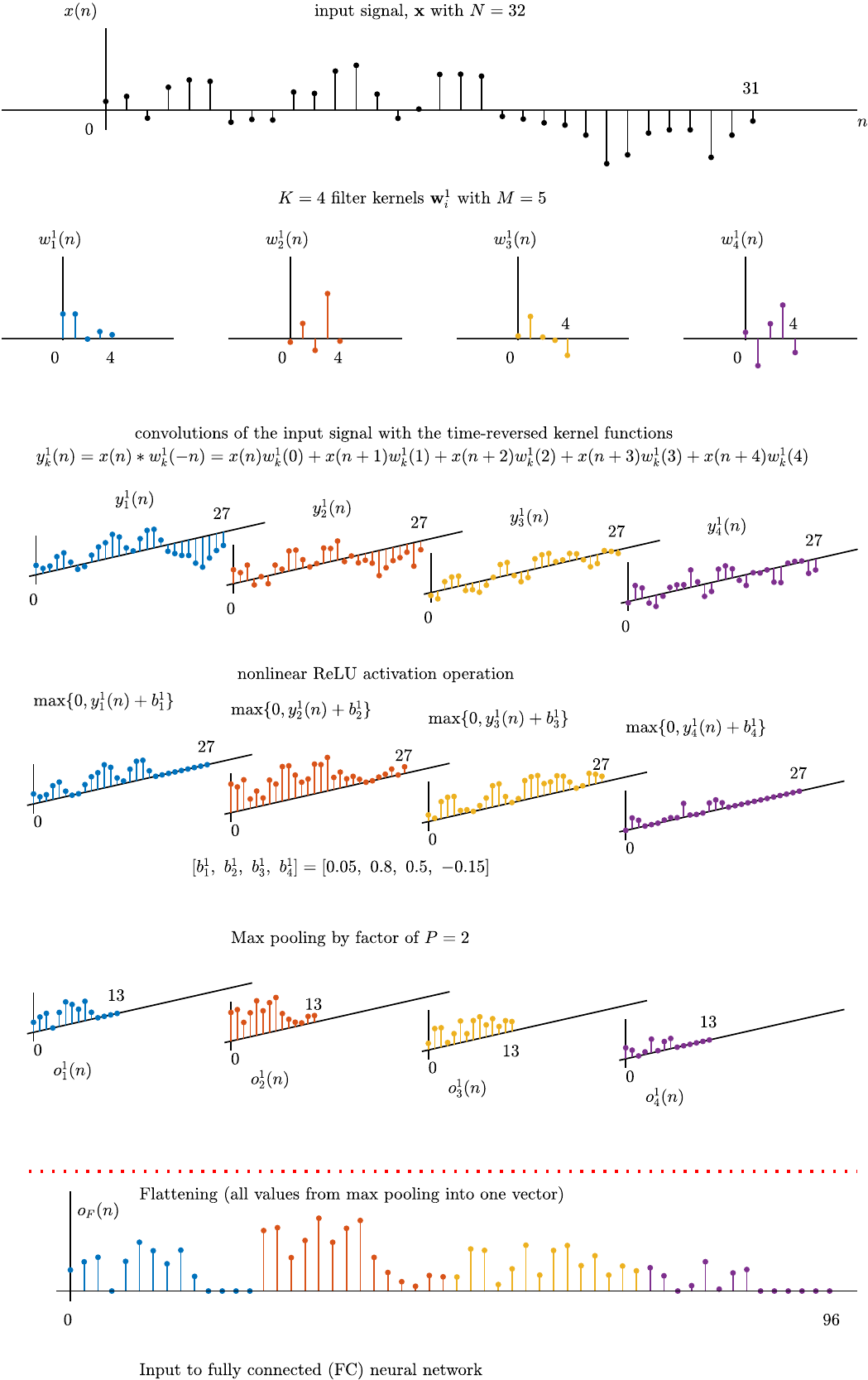}%
			\caption{Illustration of the operation of the first convolutional layer for a CNN with an input signal of $N=32$ samples each, four convolution filters ($K=4$) with $M=5$ samples, the ReLU nonlinear activation function $\max\{0,y^1_k+b^1_k\}$, $k=1,2,3,4$, and max pooling with the factor $P=2$. The output from the max pooling operation is used either  as input to the next convolution layer, or it is flattened  and presented to the neurons of a standard fully connected (FC) neural network layer.
				The weights of the convolution filter are generated as Gaussian random numbers (common way for the CNN initialization).}%
			\label{CNN_conv}% 
		\end{center}
	\end{figure}

	\bigskip
	
	\item \textbf{Repeated convolutions:}  Notice that before flattening, the convolution  steps can be repeated one or more times, involving different sets of filter functions (features). Such repeated convolutions help to find possible hierarchically composed features. The convolutional steps can be repeated with or without the activations and pooling functions,  referred to as \textit{convolution-activation-pooling}.

	\begin{exmpl}
		The output signals,   $o_1^1(n)$, $o_2^1(n)$, $o_3^1(n)$, and $o_4^1(n)$ from the first convolutional layer in Example   \ref{cnn_ex1} are used as input to the second convolutional layer of a CNN, as shown in Fig. \ref{CNN_conv2}. These signals are processed with $K=5$ convolutional filters, $w_{1,p}^2(n)$, $w_{2,p}^2(n)$, $w_{3,p}^2(n)$,  $w_{4,p}^2(n)$, and $w_{5,p}^2(n)$, each of length $M=3$ and for $p=1,2,3,4$. The output of the convolutional filters in the second layer is denoted by $y_1^2(n)$, $y_2^2(n)$, $y_3^2(n)$, $y_4^2(n)$, and $y_5^1(n)$. The ReLU activation function is applied to these signals to produce, $\max\{0,y_k^2(n)+b^2_k\}$, $k=1,2,3,4,5$. The max-pooling stage with factor of $P=2$ is next used to produce the signals  $o_1^2(n)$, $o_2^2(n)$, $o_3^2(n)$, $o_4^2(n)$ and $o_5^2(n)$. Finally, the flattened output of the second convolutional layer, denoted by $o_F^2(n)$, is formed to serve as an input to the FC layer.
	\end{exmpl}
	
	\begin{figure}[hptb]
		\begin{center}
			\includegraphics[]{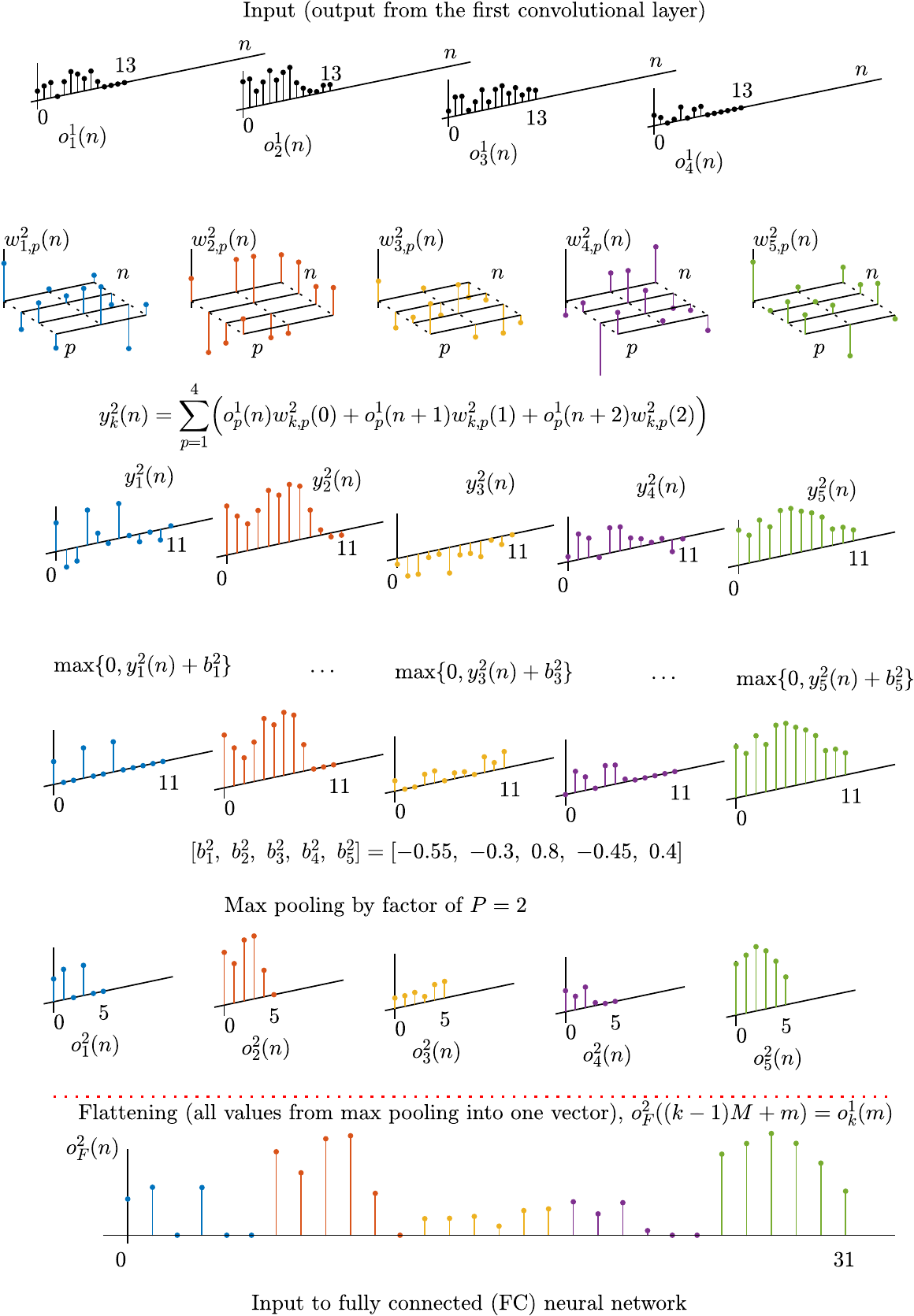}%
			\caption{Operation of the second convolution layer for a CNN which uses the output from the first layer  as its input. Five convolution filters ($K=5$) with $M=3$ samples were used, the ReLU nonlinear activation function, $\max\{0,y^2_i+b^2_i\}$, and max pooling with factor $P=2$, whereby the signal from the previous step is grouped into segments of two samples with the largest sample serving as the output. The signal from the max pooling is used either  as an input to the next convolution layer or it is flattened (if only one convolutional layer is used) and fed to fully connected (FC) of a standard neural network. The weights of the convolution filter are generated as Gaussian random numbers (common way for the CNN initialization).}%
			\label{CNN_conv2}%
		\end{center}
	\end{figure}

	\bigskip

	\item \textbf{Fully Connected (FC) Layers:} The output of the previous convolutional steps, after flattening, are connected in the form of the flattened data to the standard neural network with fully connected neurons. Neurons in this layer have full connectivity with all neurons in the preceding and following layers, as seen in regular feed-forward neural networks.  The FC layers may have a traditional multilayer structure, and are followed by the output layer, which is described in the sequel.

	\begin{figure}[hptb]
		\begin{center}
			\includegraphics{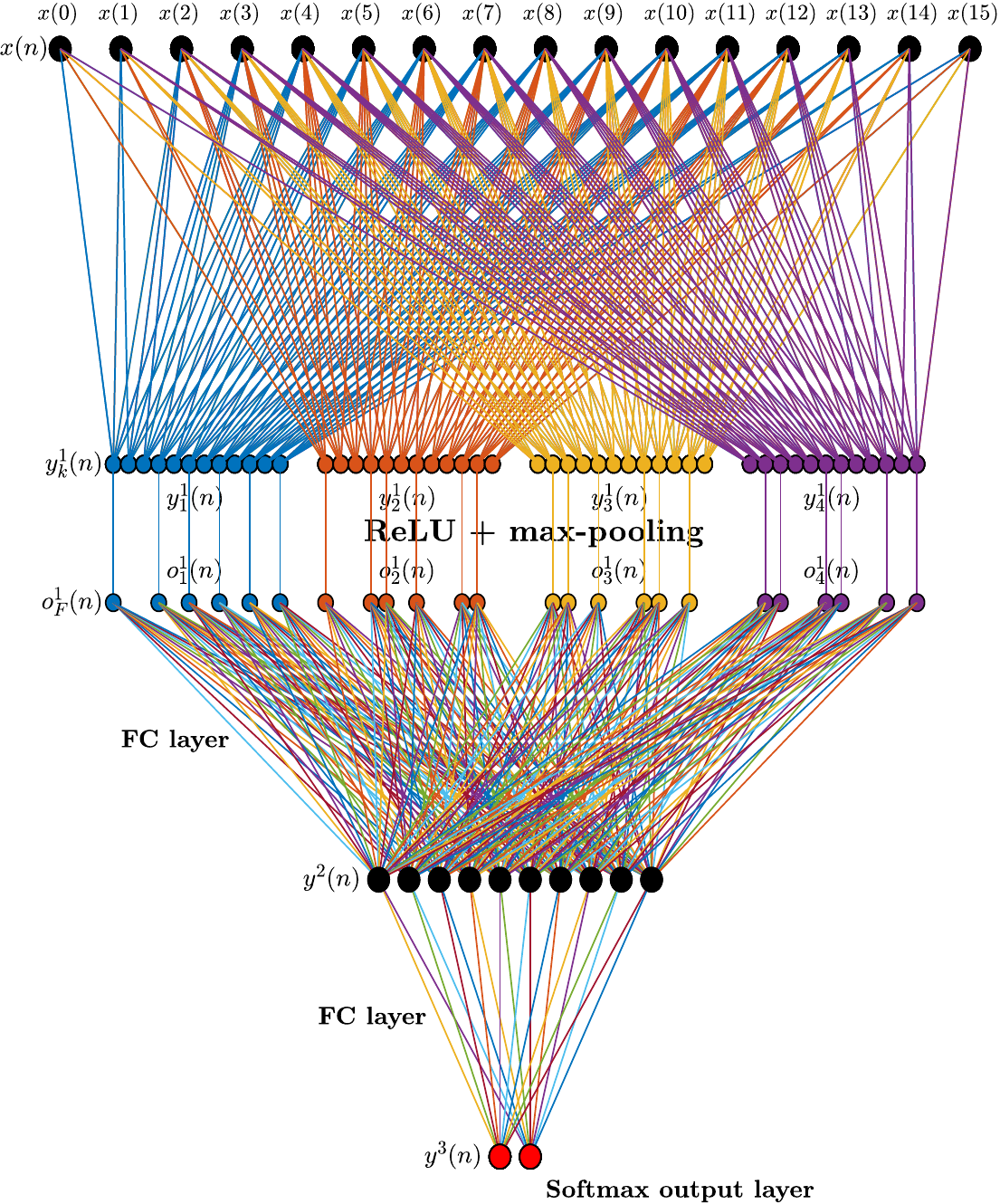} (a) \\
			\includegraphics[trim={0cm 4cm 0 3cm},clip,scale=0.40]{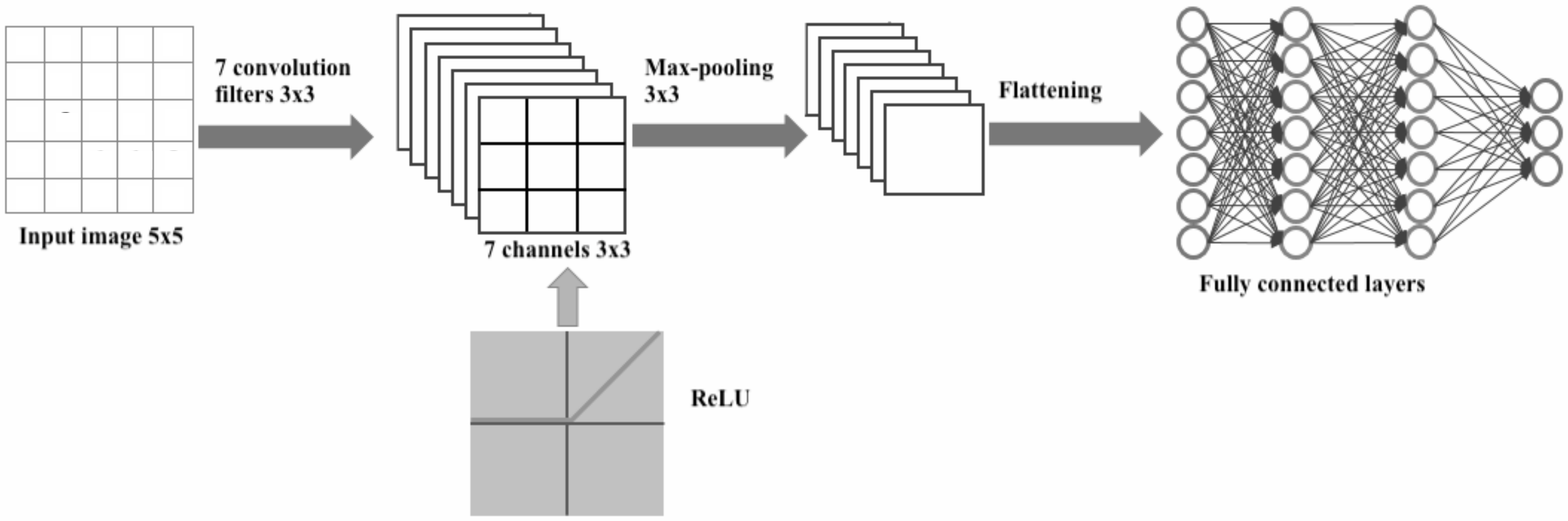} (b)%
			\caption{An exemplar of architecture of CNNs for signals and images. (a) Illustration of a CNN for signals, with one convolution layer and two FC layers, with  two neurons at the output (softmax) layer. (b) Illustration of a CNN for images with one convolution layer and two FC layers, with three neurons at the output (softmax) layer.}%
			\label{fully_connected_layer11}%
		\end{center}
	\end{figure}

\end{enumerate} 
\begin{exmpl}\label{cnn_ex11}
	An input signal, $x(n)$, with $N=32$ samples, serves as input to the one-dimensional convolutional layer in a CNN, as shown in Fig. \ref{CNN_conv} and Fig. \ref{CNN_conv2}. The signal is processed with $K=4$ convolutional filters, $w_1^1(n)$, $w_2^1(n)$, $w_3^1(n)$, and $w_4^1(n)$, each of length $M=3$. The output of these convolutional filters is  given by $y_1^1(n)$, $y_2^1(n)$, $y_3^1(n)$, and $y_4^1(n)$. The ReLU activation function is applied to these signals to produce, $\max\{0,y_1^1(n)+b_1\}$, $\max\{0,y_2^1(n)+b_2\}$, $\max\{0,y_3^1(n)+b_3\}$, and $\max\{0,y_4^1(n)+b_4\}$. The max-pooling with factor $P=2$ yields the signals  $o_1^1(n)$, $o_2^1(n)$, $o_3^1(n)$, and $o_4^1(n)$. Finally, the flattened output of this layer is formed, and denoted by $o_F^1(n)$.
\end{exmpl}

\begin{exmpl}
	To illustrate the sheer number of parameters required in one successful example of a CNN we quote the authors of  \textbf{AlexNet}:

	"We trained a large, deep convolutional neural network to classify the 1.2 million high-resolution images in the ImageNet LSVRC-2010 contest into the 1000 different classes.  The neural network, which has 60 million parameters and 650,000 neurons, consists of five convolutional layers, some of which are followed by max-pooling layers, and three fully connected layers with a final 1000-way softmax.
	
	The first convolutional layer filters the 224x224x3 input image with 96 kernels of size 11x11x3 with a stride of 4 pixels (this is the distance between the receptive field centers of neighboring neurons in a kernel map). The second convolutional layer takes as input the (response-normalized and pooled) output of the first convolutional layer and filters it with 256 kernels of size 5x5x48. The third, fourth, and fifth convolutional layers are connected to one another without any intervening pooling or normalization layers. The third convolutional layer has 384 kernels of size 3x3x256 connected to the (normalized, pooled) outputs of the second convolutional layer. The fourth convolutional layer has 384 kernels of size 3x3x192, and the fifth convolutional layer has 256 kernels of size 3x3x192. The fully-connected layers have 4096 neurons each." A. Krizhevsky, I. Sutskever, and G. E. Hinton, 
	"ImageNet Classification with Deep Convolutional Neural Networks", \textit{Communications of the ACM}, 60 (6): 84--90, May 2017. 
\end{exmpl}

\begin{figure}[hptb]
	\includegraphics[trim=2cm  2cm 2cm 2cm,clip,width=\textwidth]{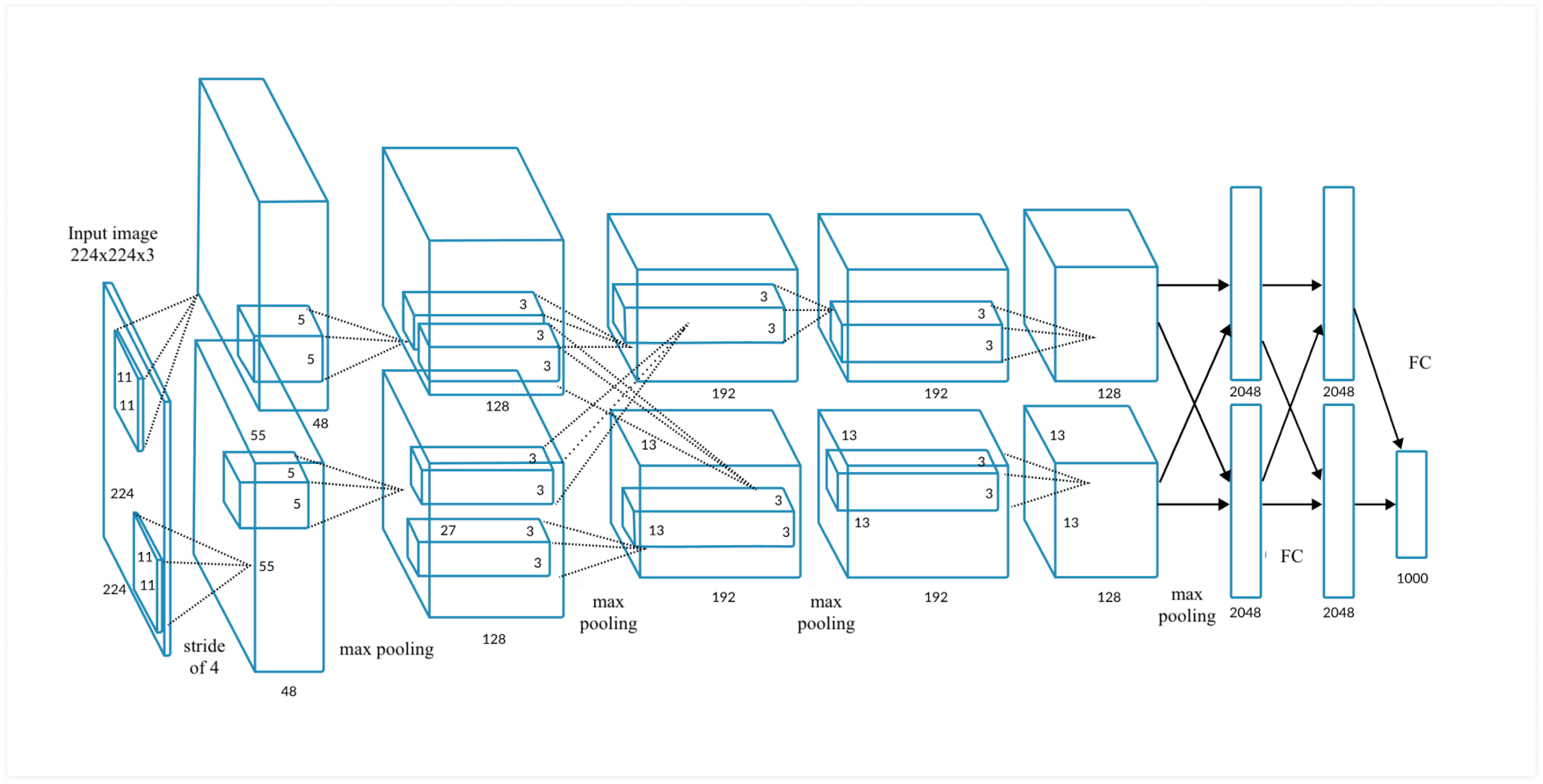}%
	
	\textbf{AlexNet} architecture \cite{krizhevsky2012imagenet}:
	
	[224x224x3] INPUT COLOR IMAGES
	
	CONV1: 96 11x11 filters at stride 4, pad 0 $-->$ 2 x [55x55x48]= [55x55x96] 
	
	MAX POOL1: 3x3 filters at stride 2 $-->$  [27x27x96]  
	
	CONV2: 256 5x5 filters at stride 1, pad 2  $-->$ [27x27x96]
	
	MAX POOL2: 3x3 filters at stride 2  $-->$ [13x13x256] 
	
	CONV3: 384 3x3 filters at stride 1, pad 1 $-->$        [13x13x256] 
	
	CONV4: 384 3x3 filters at stride 1, pad 1  $-->$ [13x13x384] 
	
	CONV5: 256 3x3 filters at stride 1, pad 1 $-->$        [13x13x256] 
	
	MAX POOL3: 3x3 filters at stride 2 $-->$     [6x6x256]       
	
	[4096] FC6: 4096 neurons
	
	[4096] FC7: 4096 neurons
	
	[1000] FC8: 1000 neurons (class scores)

	\caption{AlexNet was designed by SuperVision group (2012) \cite{krizhevsky2012imagenet}.  This deep CNN is used to classify the 1.2 million high-resolution images in the ImageNet LSVRC-2010 contest into the 1000 different classes.  The network has 60 million parameters and 650,000 neurons, consists of five convolutional layers, some of which are followed by max-pooling layers, and three fully connected layers with a final 1000-way softmax.}%
	\label{AlexNet_2012}%
\end{figure}

\section{Updating Convolution Weights: Back-propagation}\label{BackPropag}

The initial parameters (weights) of a CNN are typically updated in a supervised way through a gradient-based learning process known as the back-propagation (BP) algorithm. For each iteration of the BP, the gradient magnitude (or sensitivity) of each network parameter (such as the weights of the convolutional or the fully-connected layers) is computed. These parameter sensitivities are then used to iteratively update the CNN parameters until a certain stopping criterion is met or the training data set is exhausted. 

\subsubsection{Initialization}
 Unlike standard adaptive systems, where the initial weight values are typically   set to zero, in neural networks the initial values of the weights are typically assumed as random (and different) values for each channel and layer.  Since the weights $w_k(m)$ multiply, in general, $N_{in}$ input signal values (at the considered input neurons of the layer), the only requirement is that the choice of the initial weights preserves the expected energy of the output for the considered layers. This  is achieved, for example,  if the initial weights are \textit{Gaussian distributed}, with 
$$w_k(m) \sim \sqrt{\frac{2}{N_{in}}}\mathcal{N}(0,1).$$
The factor of 2 is used since the ReLU activation function will remove negative output values, which accounts for a half of the expected energy. 

Another possibility is to use \textit{uniformly} distributed initial wights, $w_k(m)$, whereby the sum of $N_{in}$ initial weights, $\sum_{m=0}^{N_{im}} w_k(m)$, produces a unit variance. Such uniformly distributed weights are defined by the interval
$$w_k(m) \sim \Big[-\sqrt{\frac{6}{N_{in}}}, \ \sqrt{\frac{6}{N_{in}}}\Big].$$
The variance of this random variable is $\mathrm{Var}\{w_k(m)\}=\frac{6}{N_{in}}\frac{1}{3}$. The variance of a sum of $N_{it}$ values, divided by 2, to take into account the ReLU, produces unit variance. Such initial values are called the He initial values. 

If the previous values of the initial weights are additionally reduced, taking into account the number of output neurons for the considered layer, $N_{out}$, then the Xavier  initial values are obtained 
$$w_k(m) \sim \sqrt{\frac{2}{N_{in}+N_{out}}}\mathcal{N}(0,1)$$
or
$$w_k(m) \sim \Big[-\sqrt{\frac{6}{N_{in}+N_{out}}}, \ \sqrt{\frac{6}{N_{in}+N_{out}}}\Big].$$

\subsubsection{Back-propagation in a two-layer CNN}
Consider first the weight update in the simplest CNN which consists of two layers, a convolutional layer and a fully connected output layer. 

\noindent\textbf{Convolutional layer.} For the input  $\mathbf{x}=[x(0), \ x(1), \dots, x(N-1)]^T$, the output signal of the convolutional layer of the CNN, with $K$ filters of the width $M$, is given by
\begin{equation}y_k^1(n)=w_k^1(0)x(n)+w_k^1(1)x(n+1)+\dots+w_k^1(M-1)x(n+M-1)=\sum_{m=0}^{M-1}w_k^1(m)x(n+m),
\label{ConvCNNK}
\end{equation}
for the channels $k=1,2,\dots,K$, as shown in Fig. \ref{CNN_conv}. The overall output of the convolution layer is then obtained after the bias term is included and upon the application of the ReLU activation function, to yield
\begin{equation}o_k^1(n)=f(y_k^1(n)+b_k^1).
\label{ConvCNN)}
\end{equation}
For simplicity, we shall first assume that no max-pooling or any other down-sampling is performed. 

The output from the convolutional layer is then stacked into a vector of length $K(N-M+1)$, which serves as input to the fully connected layer with  $S$ outputs.

Each of $K(N-M+1)$ nodes of the output of the convolutional layer, with the signal samples, 
$$[o_1^1(0),\dots,o_1^1(N-M), \ o_2^1(0),\dots,o_2^1(N-M), \dots, o_K^1(0),\dots,o_K^1(N-M)]^T$$ 
is connected  to each of the $S$ nodes of the fully connected output layer to produce the overall CNN output of the form
\begin{align}y_k^2 & =w_{k}^2(0)o_1^1(0)+w_{k}^2(1)o_1^1(1)+\dots+w^2_{k}(N-M)o_1^1(N-M) \nonumber  \\& +w_{k}^2(N-M+1)o_2^1(0)+\dots+w^2_{k}(2(N-M+1)-1)o_2^1(N-M) \nonumber  \\
& \vdots  \nonumber  \\
& +w_{k}^2((K-1)(N-M+1))o_K^1(0)+\dots+w^2_{k}(K(N-M+1)-1)o_K^1(N-M),
\label{ConvCNNFC}
\end{align}
for $k=1,2,\dots,S$. Note that the number of weights in the $k$th FC layer is $SK(N-M+1)$.

A commonly used \textit{loss function} in the minimization  is the mean square error (MSE) between the network prediction and the true lable, given by  
\begin{equation}\mathcal{L}=\frac{1}{2}\sum_{k=1}^S (y^{2}_k-t_k)^2,
\label{lossFL}
\end{equation}
where $t_k$ is the desired or target output (also called a teaching signal). 

\textbf{Training process.} To define the gradient descent relations for the update of all previous weights (within the convolutional layer and the fully connected layer) in the training process, 
consider first the convolutional layer, in (\ref{ConvCNNK})-(\ref{ConvCNN)}), to give the gradient weight update in the form
\begin{equation}w^1_{k}(m)_{new}= w^l_{k}(m)_{old}-\alpha \frac{\partial \mathcal{L}}{\partial w^1_{k}(m)} \ _{{|w^1_{k}(m)=w^1_{k}(m)}_{old}}.\label{GradUpCNN2L}\end{equation}
The element-wise gradient values are then calculated
\begin{equation}\frac{\partial \mathcal{L}}{\partial w^1_{k}(m)}=\sum_n \frac{\partial \mathcal{L}}{{\color{red}\partial y^1_{k}(n)}}\frac{{\color{red}\partial y^1_{k}(n)}}{\partial w^1_{k}(m)}=\sum_n \frac{\partial \mathcal{L}}{\partial y^1_{k}(n)}x(n+m)= \frac{\partial \mathcal{L}}{\partial y^1_{k}(m)}*_cx(m), \label{GradUpCNNGV}\end{equation}
where (\ref{ConvCNNK}) is used for the calculation\footnote{Here, we have also used the property of an implicit function derivative, given by \begin{gather*}
		\frac{\partial F(u(x,y,z),v(x,y,z),w(x,y,z))}{\partial x}=\frac{\partial F(u(x,y,z),v(x,y,z),w(x,y,z))}{\partial u}  \frac{\partial u}{\partial x} \\ 
		+\frac{\partial F(u(x,y,z),v(x,y,z),w(x,y,z))}{\partial v}  \frac{\partial v}{\partial x}+\frac{\partial F(u(x,y,z),v(x,y,z),w(x,y,z))}{\partial w}  \frac{\partial w}{\partial x}.\end{gather*}} of $\partial y^1_{k}(n)/\partial w^1_{k}(m)$.

Next, we need  to calculate the so called \textit{delta error} function $\partial \mathcal{L}/\partial y^1_{k}(m)=\Delta_k^1(m)$, which can be written as
\begin{gather}\Delta_k^1(m)=\frac{\partial \mathcal{L}}{\partial y^1_{k}(m)}=\sum_p \frac{\partial \mathcal{L}}{{\color{red} \partial y^2_{p}}}\frac{{\color{red} \partial  y^2_{p}}}{\partial y^1_{k}(m)}=\sum_p \frac{\partial \mathcal{L}}{\partial y^2_{p}}\frac{\partial  y^2_{p}}{{\color{red}\partial o^1_{k}(m)}}\frac{{\color{red}\partial  o^1_{k}(m)}}{\partial y^1_{k}(m)} \nonumber \\
=\sum_p \Delta_p^2w^2_{p}((k-1)M+m) \ \ u(y^1_k(m)) \label{GradUpCNNGBP11}
\end{gather}
where the relation in (\ref{ConvCNNFC}) is used for the calculation of $\partial  y^2_{p}/\partial o^1_{k}(m)=w^2_{p}((k-1)M+m)$ and 
$$\Delta_p^2=\frac{\partial \mathcal{L}}{\partial y^2_{p}}= y^{2}_p-t_p$$ is the error in the final stage. 

The relation in (\ref{GradUpCNNGBP11}) back-propagates the error from layer 2, denoted by $\Delta_p^2$, to layer 1, to yield a portion of the overall error attributed to neuron $k$ of layer 1, $\Delta_k^1(m)$. We can now calculate $\partial \mathcal{L}/\partial y^1_{k}(m)=\Delta_k^1(m)$ and the gradient for the update in (\ref{GradUpCNN2L}).

The bias terms are updated in the same way 
\begin{equation}b^1_{k,new}=b^1_{k,old}-\alpha \frac{\partial \mathcal{L}}
{\partial b^{1}_k} \ _{|b_k^1=b_{k,old}^1} \label{BGradUpCNN22}
\end{equation}
and
\begin{equation}\frac{\partial \mathcal{L}}{\partial b^1_{k}}=\sum_n \frac{\partial \mathcal{L}}{{\color{red}\partial y^1_{k}(n)}}\frac{{\color{red}\partial y^1_{k}(n)}}{\partial b^1_{k}}=\sum_n \frac{\partial \mathcal{L}}{\partial y^1_{k}(n)}=\sum_n \Delta_k^1(n) \label{GradUpCNNB}\end{equation}

If the max-pooling operation is used, then the output $y_k(n)$ is used only for some $n\in \mathbb{M}_k$, and the gradient update is adjusted accordingly, as 
\begin{equation}\frac{\partial \mathcal{L}}{\partial w^1_{k}(m)}=\sum_{n \in \mathbb{M}_k}  \frac{\partial \mathcal{L}}{\partial y^1_{k}(n)}\frac{\partial y^1_{k}(n)}{\partial w^1_{k}(m)}=\sum_{n \in \mathbb{M}_k} \frac{\partial \mathcal{L}}{\partial y^1_{k}(n)}x(n+m). \label{GradUpCNNGMP}\end{equation}

Notice that within the max-pooling, the convolution values used at $n\in \mathbb{M}_k$ may change at each update step. If the stride is also used, then the values of $y_k(n)$ are calculated according to a defined stride step. For example, with the stride value of $2$, the convolutions are always calculated at $y_k(0)$, $y_k(2)$, $\dots$, $y_k(N-2)$.

\noindent\textbf{ Fully Connected (FC) layer.}  The input to the FC layer represents the flattened output from the convolutional layer, given by
$$o_F^1((k-1)(N-M+1)+m)=o^1_k(n).$$
The indices $n$ in $o_F^1(n)$ range from $0$ to $K(N-M+1)-1$. 
Notice that relation (\ref{ConvCNNFC}) could be equally written as 
$$y_k^2=\sum_{n=0}^{K(N-M+1)-1}w^2_{k}(n)o^1_F(n).$$

The update of the fully connected layer weights, $w^2_{k}(n)$, is performed in the same way, using
\begin{equation}w^2_{k}(m)_{new}= w^2_{k}(m)_{old}-\alpha \frac{\partial \mathcal{L}}{\partial w^2_{k}(m)} \ _{{{|w^2_{k}(m)=w^2_{k}(m)_{old}}}},\label{GradUpCNN2FC}\end{equation}
with the gradient elements in the form
$$\frac{\partial \mathcal{L}}{\partial w^2_{k}(m)} = \frac{\partial \mathcal{L}}{{\color{red}\partial y^2_{k}}} \frac{{\color{red}\partial y^2_{k}}}{\partial w^2_{k}(m)}=(y^{2}_k-t_k)o^1_F(m).$$
Notice that this relation is \textit{a special case} of (\ref{GradUpCNNGMP}), for the CNN with $y_k^2(n)=y_k^2(0)=y_k^2$, that is, when the summation over $n$ in  (\ref{GradUpCNNGMP}) reduces to one term only for $n=0$. 

If a nonlinear activation function is used at the output, then the factor of $f'(y^2_k)$ should multiply the right hand side of $\partial \mathcal{L}/\partial w^2_{k}(m)$.

\subsubsection{ Softmax Output Layer} 

In some applications, the output layer gives the probabilities for the decision when classifying of the analyzed data. The output therefore represents a list of probabilities for different possible labels (basins of attraction) associated with the analyzed signal or image (for example, dog, cat, bird in the image), whereby  the label that receives the highest probability is the classification decision. In the error calculation, the desired (target) output then assumed the value $t_k=1$ for one value  $k=k_0$ (in training process we know what signal/image is analyzed by the CNN) and $t_k=0$ for other values of $k$.

Since the output, $y^L_k$, from the last $L$th layer (overall output), may assume various positive and negative real values, we need to map the output $y^L_k$ into probability-like values, using a function of the form
\begin{equation}P_k=\frac{e^{y^L_k}}{\sum_{i=1}^Se^{y^L_i}}.\label{softmaxM}\end{equation}
called the softmax. Obviously, $0 \le P_k \le 1$ and $\sum_{k=1}^SP_k=1$. 

When the softmax is used as the output mapping, the loss function is modified accordingly, from the mean square error to the 
 cross-entropy form, given by
$$\mathcal{L}=-\sum_{k=1}^St_k\ln(P_k).$$
This cross-entropy is very large if there is a $t_k$ close to $1$, but the corresponding output probability $P_k$ is small, meaning that a big change in the weights should be performed. The cross-entropy, $\mathcal{L}$, is small only when for $t_{k_0}=1$ at a specific $k_0$, and the value of corresponding $P_{k_0}$ is close to $1$. 

We can easily show that the delta error function in the  output layer is of the form 
$$\Delta^L_k=\frac{\partial \mathcal{L}}{\partial y^L_k}=\sum_{i=1}^S\frac{\partial \mathcal{L}}{{\color{red}\partial P_i}}\frac{{\color{red}\partial P_i}}{\partial y^L_k}=\sum_{i=1}^S\Big(\frac{t_i}{P_i}P_iP_k\Big)-\frac{t_k}{P_k}P_k=P_k-t_k$$
since from (\ref{softmaxM}) it follows that $\partial P_i /\partial y^L_k=-P_iP_k$ if $i \ne k$  and $\partial P_i /\partial y^L_k=P_i(1-P_k)=-P_iP_k+P_k$ if $i=k$, while $\sum_{i=1}^St_i=1$. 

Therefore, as expected, there is no weight correction if $t_k=P_k$, while, as desired, all the previous (and next) relations regarding the back-propagation also hold in this case.

\subsubsection{Back-Propagation in a Multi-Layer CNN}
After the back-propagation is illustrated for a simple two-layer network example, we can now generalize the back-propagation relations to a multi-layer CNN. The output in the layer $l$, $l=1,2,\dots,L$, of a general CNN without max-pooling, is defined  by  
$$\mathbf{y}_k^l=\sum_p \mathbf{o}_p^{l-1}*_c\mathbf{w}_{k,p}^l+b_k^l,$$
where $\mathbf{o}^{l-1}$ is the output of the layer $(l-1)$, as shown in Fig. \ref{CNN_conv2}. The element-wise form  of this output is given by
\begin{equation}y^l_k(n)= \sum_{p=1}^K\sum_{m=0}^{M-1}\Big(o_p^{l-1}(n+m)w^{l}_{k,p}(m)\Big)+b_k^l. \label{inOutML}
\end{equation}
Notice that the input to the layer input is equal to the input signal, $\mathbf{o}^{0}=\mathbf{x}$. For any other layer we have 
$$\mathbf{o}^{l-1}=f(\mathbf{y}_k^{(l-1)}),$$
where $f(x)$ is the nonlinear activation function (commonly ReLU in the CNN). 

Next, we specify those derivatives in (\ref{inOutML}) that will be used in the update of neural network weights
\begin{align*} & \frac{\partial y_k^l(n)}{\partial w^l_{k,p}(m)}  =o_p^{(l-1)}(n+m)
\\  &\frac{\partial y_k^{l+1}(n-\mu)}{\partial o_q^{l}(n) }  = \frac{\partial }{\partial o_q^{l}(n) } \Big( \sum_{p=1}^K\sum_{m=0}^{M-1}\Big(o_p^{l}(n-\mu+m)w^{l+1}_{k,p}(m)\Big)+b_k^{l+1}\Big)=w^{l+1}_{k,q}(\mu)
\\
& \frac{\partial o_p^{l}(n)}{\partial y_k^{l}(n)}  =f'(y_k^{l}(n))=u(y_k^{l}(n)),
\end{align*}
where $u(x)$ is the unit step function.

\medskip

\noindent\textbf{Gradients of weight update.} The weights should be changed according to the gradient descent direction of the loss function, $\mathcal{L}$, that is
\begin{equation}w^l_{k,p}(m)_{new}= w^l_{k,p}(m)_{old}-\alpha \frac{\partial \mathcal{L}}{\partial w^l_{k,p}(m)} \ _{{|w^l_{k,p}(m)=w^l_{k,p}(m)}_{old}}.\label{GradUpCNN}\end{equation}

\begin{itemize}  
	\item      
	For the convolutional layer, the derivative of the cost function with respect to $ w^l_{k,p}(m)$, using the previously stated derivatives, becomes
	\begin{gather*}\frac{\partial \mathcal{L}}{\partial w_{k,p}^l(m)}=\sum_n \frac{\partial \mathcal{L}}{{\color{red}\partial y_k^l(n)}}\frac{{\color{red}\partial y_k^l(n)}}{\partial w_{k,p}^l(m)}=\sum_n \frac{\partial \mathcal{L}}{\partial y_k^l(n)}o_p^{(l-1)}(n+m)
	=\frac{\partial \mathcal{L}}{\partial y_k^l(m)}*_co_p^{(l-1)}(m).\end{gather*}
	
	\item
	For the standard, fully connected layer, according to (\ref{ConvCNN3N1}), the following holds
	$$\frac{\partial \mathcal{L}}{\partial w_k^l(m)}= \frac{\partial \mathcal{L}}{{\color{red}\partial y_k^l(0)}}\frac{{\color{red}\partial y^l_k(0)}}{\partial w_k^l(m)}=\frac{\partial \mathcal{L}}{\partial y_k^l}\frac{\partial y_k^l}{\partial w_k^l(m)}= \frac{\partial \mathcal{L}}{\partial y_k^l}o_k^{(l-1)}(m).$$
\end{itemize}

\subsubsection{Delta error back-propagation}        
In the CNN jargon, the derivative  $\partial \mathcal{L}/\partial y_k^l(m)=\Delta_k^{l}(m)$ is called the \textit{delta error}. For an arbitrary layer $l$, it should be related to the error function in the last (output) layer, $\Delta^{L}_k=P_k-t_k$. By using the composition of derivatives, we can relate the delta error in the $l$th layer with that in the next, $(l+1)$th, layer, and then propagate this relation iteratively to the output layer. This can be written as  
%
%$$\Delta_k^l(m)=\frac{\partial \mathcal{L}}{\partial y_k^l(m)}=\frac{\partial \mathcal{L}}{\partial y_p^L} \ \frac{\partial y_p^{L}}{\partial y_q^{L-1}(m)} \ \frac{\partial y_q^{L-1}(m)}{\partial y_r^{L-2}(m)}\cdots \frac{\partial y_s^{l+2}(m)}{\partial y_t^{l+1}(m)}\frac{\partial y_t^{l+1}(m)}{\partial y_k^{l}(m)}$$   
% Since 
% $$o_k^{l}(m)=f(y_k^{l}(m))=\max\{ 0,y_k^{l}(m) \}$$
%we get
%$$ \frac{\partial y_k^{l}(m)}{\partial y_p^{l-1}(m)}= \sum_{n} \frac{\partial y_k^{l}(m)}{\partial o_p^{l-1}(m+n)} \frac{\partial o_p^{l-1}(m+n)}{\partial y_p^{l-1}(m)}=\sum_nw_{k,p}^l(n)u(y_p^{l-1}(m)),???$$
%where $u(y)$ is the unite step function.
%       
\begin{gather*}\Delta_k^l(n)=\frac{\partial \mathcal{L}}{\partial y_k^l(n)}=
\sum_{m}  \sum_{p} 
\frac{\partial \mathcal{L}}{{\color{red}\partial y_p^{l+1}(n-m)}}  \frac{{\color{red}\partial y_p^{l+1}(n-m)}}{\partial y_k^{l}(n)} \\
=  \sum_{m}  \sum_{p} 
\frac{\partial \mathcal{L}}{\partial y_p^{l+1}(n-m)}  \frac{\partial y_p^{l+1}(n-m)}{{\color{red}\partial o_k^{l}(n) }}  \frac{{\color{red}\partial o_k^{l}(n)}}{\partial y_k^{l}(n)}=
\sum_{m} \sum_{p} 
\frac{\partial \mathcal{L}}{\partial y_p^{l+1}(n-m)} w^{l+1}_{p,k}(m) \ \ u(y^l_k(n)).
\end{gather*}

\medskip

\noindent\textbf{Back-propagation of the delta error.} 
From the above, the recursive back-propagation relation for the delta error calculation in the convolutional layer is given by
\begin{gather*}
\Delta_k^l(n)= u(y^l_k(n))
\sum_{p}  \Big(\sum_{m} 
\frac{\partial \mathcal{L}}{\partial y_p^{l+1}(n-m)} w^{l+1}_{p,k}(m)\Big) =u(y^l_k(n))\sum_{p} \Big(
\Delta_p^{l+1}(n) * w^{l+1}_{k,p}(n) \Big),
\end{gather*}
with the final value (the initial value for the back-propagation) for the mean square error
$$\Delta_k^{L}=\frac{\partial \mathcal{L}}{\partial y_k^L}=\frac{1}{\partial y_k^L} \Big(\frac{1}{2}\sum_p(y_p^L-t_p)^2\Big)=y_k^L-t_k,$$
while for the Softmax layer we have
$$\Delta^{L}_k=P_k-t_k$$
for the Softmax layer.

For a fully connected layer, in the standard neural network, we obtain 
\begin{gather*}
\Delta_k^l=\frac{\partial \mathcal{L}}{\partial y_k^l}=
\sum_{p} 
\frac{\partial \mathcal{L}}{\partial y_p^{l+1}}  \frac{\partial y_p^{l+1}}{\partial y_k^{l}}=
\sum_{p} 
\frac{\partial \mathcal{L}}{\partial y_p^{l+1}}  \frac{\partial y_p^{l+1}}{\partial o^{l}(k)}\frac{\partial o^{l}(k)}{\partial y_k^{l}}= \sum_{p}
\Delta_p^{l+1}w^{l+1}_{p}(k) \ \ u(y_k^l).
\end{gather*}

\noindent\textbf{Bias update.} The bias propagation obeys similar rules, and is given by
$$\frac{\partial \mathcal{L}}{\partial b_k^l}=\sum_n \frac{\partial \mathcal{L}}{\partial y_k^l(n)}\frac{\partial y_k^l(n)}{\partial b_k^l}=\sum_n \frac{\partial \mathcal{L}}{\partial y_k^l(n)}=\sum_n \Delta_k^l(n),$$
with
\begin{equation}b^l_{k,new}=b^l_{k,old}-\alpha \frac{\partial \mathcal{L}}
{\partial b^{l}_k} \ _{| b^{l}_k=b^l_{k,old}} \label{BGradUpCNN}
\end{equation}

For the FC layers, the bias update is performed according to
$$\frac{\partial \mathcal{L}}{\partial b_k^l}= \frac{\partial \mathcal{L}}{\partial y_k^l}\frac{\partial y_k^l}{\partial b_k^l}= \frac{\partial \mathcal{L}}{\partial y_k^l}= \Delta_k^l,$$
with the same update relation as in (\ref{BGradUpCNN}).

\begin{figure}[hptb]
	\begin{center}
		\includegraphics{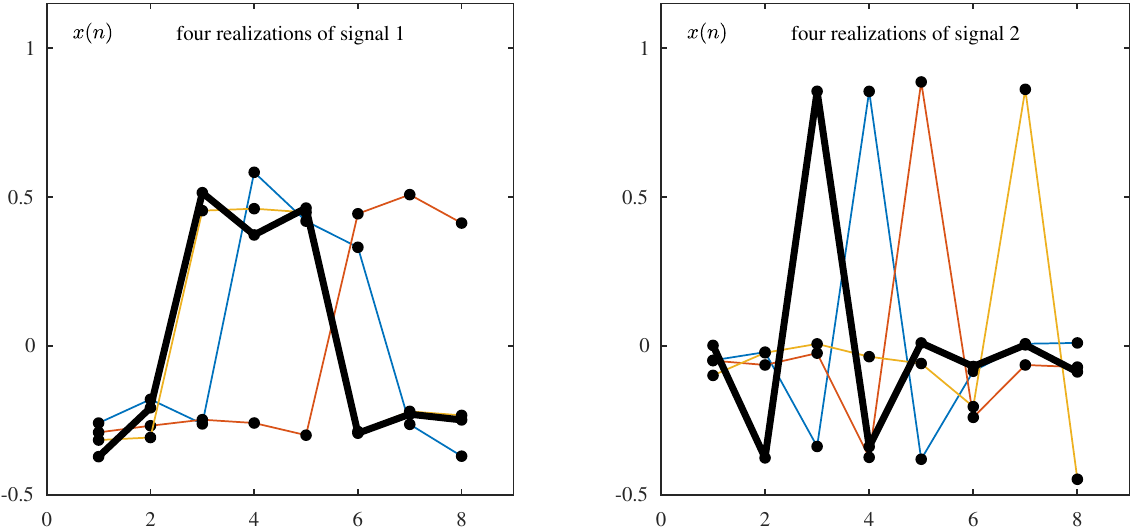} (a) \\
		\includegraphics{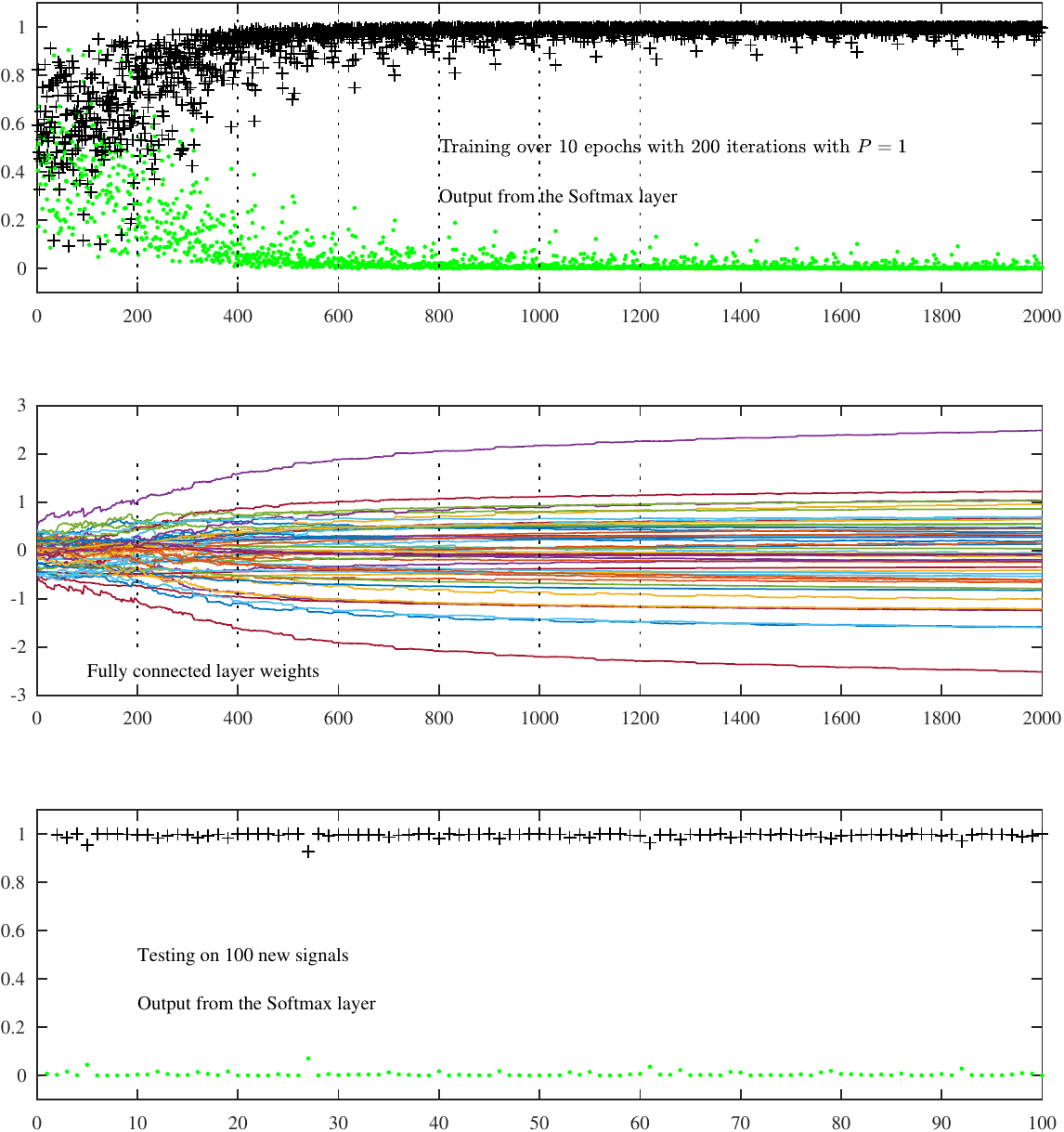} (b)%
		\caption{Operation of a CNN similar to that shown in Fig. \ref{fully_connected_layer11}, with one convolution layer and one FC layer, with two neurons at the output (softmax) layer, max-pooling with $P=1$ (no max-pooling) and $K=4$ channels.  The FC layer had therefore $((N-M+1)K) \times 2 = 24\times2=48$ weights. (a) Several random realizations of the input signal are used for the CNN training and testing. (b) Evaluation of the network output over the training process. (c) Evaluation of the FC layer weights over the training process. (d) Output of the network in the testing stage. Observe the very accurate operation of the trained CNN}%
		\label{Example_CNN}%
	\end{center}
\end{figure}

\begin{exmpl} This example of a two-layer neural network (one convolutional layer and one fully connected layer) illustrates the back-propagation operation in a step-by-step manner. The considered input signal has $N=8$ samples, which may contain either a variant of the triangular shape pattern, $feature_1=[-0.5, \ \  1, \ \  -0.5]+\nu(n)$, ($\mathbf{t}=[0, \ \ 1]^T$) or a variant of rectangular three-sample  $feature_2=[1, \ \ 1, \ \ 1]+\nu(n)$, $\mathbf{t}=[1, \ \ 0]^T$, where $\nu(n)$ is random uniform noise whose values lie in the region $0$ to $0.3$, that introduces deviations in the feature forms. The signal is embedded in additive random Gaussian noise with standard deviation of $0.05$, and then normalized to  unit energy, as shown in Fig. \ref{Example_CNN}(a).  Convolutional filters of $M=3$ samples are used to produce $K=3$ channels at the convolutional layer. The Softmax is used at the output of the FC layer, with two values that correspond to the two patterns in the target signal, $\mathbf{t}$. The network was trained using $200$ random signal realizations over 10 epochs (presented 10 times to the network). After the training, the network was tested on 100 new random signal realizations.

	\begin{centering}
		\noindent \begin{tabular}{|l|}
			\hline
			\textbf{Forward calculation}: From the input signal to the output  \\
			\hline
			$\bullet$\textbf{ Input} signal, $\mathbf{x}$, of length $N=8$,
			$\mathbf{x}=\begin{bmatrix}
			-0.18 & -0.28 & -0.23 & -0.32 & 0.45 & 0.45 & 0.45 & -0.35 \end{bmatrix}^T$. \\ The target signal was $\mathbf{t}=\begin{bmatrix}1 & 0\end{bmatrix}^T$, since $feature_2$ was present in the input. \\  
			\\
			\hline
			$\bullet$\textbf{ Weight initialization}: Random $w_k^1(m) \sim \mathcal{N}(0,1)\sqrt{2/3}$, $M=3$, for $K=3$ channels: \\
			$\mathbf{w}^1_1=\begin{bmatrix}
			-0.07 & -0.01 & -1.47 \end{bmatrix}^T$,\\
			$\mathbf{w}^1_2=\begin{bmatrix}
			\phantom{-}0.44 & \phantom{-}0.14 & -0.30 \end{bmatrix}^T$,  \\
			$\mathbf{w}^1_3=\begin{bmatrix}
			\phantom{-}1.15 & -1.01 & -1.83 \end{bmatrix}^T$.  \\
			\hline
			$\bullet$\textbf{ Convolutions}:  $\mathbf{y}_k=\mathbf{x}*_c\mathbf{w}_k^1+b_k, \ \  \ k=1,2,3$  with the initial bias values $b_1=0$, \ $b_2=0$, and $b_3=0$. \\ 
			$\mathbf{y}_1=\begin{bmatrix}
			\phantom{-}0.35 & \phantom{-}0.49 & -0.65 & -0.65 & -0.69 & \phantom{-}0.48 \end{bmatrix}^T
			$, \\
			$\mathbf{y}_2=\begin{bmatrix}
			-0.05 & -0.06 & -0.28 & -0.21 & \phantom{-}0.13 & \phantom{-}0.37 \end{bmatrix}^T
			$,  \\
			$\mathbf{y}_3=\begin{bmatrix}
			\phantom{-}0.48 & \phantom{-}0.50 & -0.77 & -1.66 & -0.76 & \phantom{-}0.71 \end{bmatrix}^T$.
			\\
			\hline
			$\bullet$\textbf{Nonlinear activation function}:  \textbf{ ReLU} activation function, $\mathbf{F}_k=f(\mathbf{y}_k^1)=\max\{0,\mathbf{y}_k^1\}$, was used, to give \\	
			$f(\mathbf{y}_1)=\begin{bmatrix}
			0.35 & 0.49 & \textbf{0.00} & \textbf{0.00} & \textbf{0.00} & 0.48 \end{bmatrix}^T$, 
			\\
			$f(\mathbf{y}_2)=\begin{bmatrix}
			\textbf{0.00} & \textbf{0.00} & \textbf{0.00} & \textbf{0.00} & 0.13 & 0.37 \end{bmatrix}^T$,  \\
			$f(\mathbf{y}_3)=\begin{bmatrix}
			0.48 & 0.50 & \textbf{0.00} & \textbf{0.00} & \textbf{0.00} & 0.71 \end{bmatrix}^T$.   \\
			\hline
			$\bullet$\textbf{ Max-pooling}: This yields the output  $o^1_k(m)=\max\{F_k(mP),\dots,F_k(mP+P-1)\}$, with $P=3$,  \\
			$\mathbf{o}^1=\begin{bmatrix} \max\{	0.35 & 0.49 & \textbf{0.00}\} & \max\{ \textbf{0.00} & \textbf{0.00} & 0.48 \} \\
			\max\{	\textbf{0.00} & \textbf{0.00} & \textbf{0.00} \} & \max\{ \textbf{0.00} & 0.13 & 0.37 \} \\
			\max\{ 0.48 & 0.50 & \textbf{0.00} \} & \max\{\textbf{0.00} & \textbf{0.00} & 0.71\}\end{bmatrix}^T=\begin{bmatrix}
			0.49 &0.48 \\
			\textbf{0.00} &0.37 \\
			0.50 &0.71 
			\end{bmatrix}^T$,  \\ 
			The indicator matrix of chosen value from ReLU, $\mathbf{M}^{ReLU}$, and max-pooling, $\mathbf{M}^{MP}$, \\
			$\mathbf{M}^{ReLU}=\arraycolsep3pt\begin{bmatrix}
			\textbf{1 }  &\textbf{1 } &0  &0  &0  &\textbf{1 } \\
			{\color{green}\textbf{0  }}&0  &0  &0  &\textbf{1 }  &\textbf{1 } \\
			\textbf{1 }  &\textbf{1 } &0 &0  &0  &\textbf{1  }
			\end{bmatrix}^T$ \ \ \ \  and \ \ \ \ $\mathbf{M}^{MP}=\arraycolsep3pt\begin{bmatrix}
			0 &\textbf{1} &0 &0 &0  &\textbf{1 } \\
			{\color{green}\textbf{1  }}&0  &0  &0  &0  &\textbf{1 } \\
			0  &\textbf{1 } &0  &0  &0  &\textbf{1  }
			\end{bmatrix}^T$,  \\
			will be used to reposition the gradient update calculated with the downsampled, $\mathbf{o}^1$, to the proper, $\mathbf{y}^1$, positions, \\ taking into account possible zeroing by the ReLU, survived from the max-pooling.  \\
			\hline
			$\bullet$\textbf{ Flattening}: $N_F=(N-M+1)P=2$, \ \  $o^1_F((k-1)N_F+m)=o^1_k(m)$, \ \ $k=1,2,3, \ \ m=0,1$. \\
			$\mathbf{o}^1_F=\begin{bmatrix} 0.49 & 0.48 & 0.00 & 0.37 & 0.50 & 0.71\end{bmatrix} ^T$, \\
			\hline
			$\bullet$\textbf{ Weight initialization}: For the FC layer, random $w_k^2(n) \sim \mathcal{N}(0,1)\sqrt{2/6}$ \\
			$\mathbf{w}^2=\begin{bmatrix}
			-0.47 &-0.06 &-0.05 &-0.81 &\phantom{-}0.62 &-0.18 \\
			\phantom{-}0.02 &\phantom{-}0.12 &-0.15 &-0.07 &-0.87 &-0.53 
			\end{bmatrix}^T$,  \\
			\hline
			$\bullet$\textbf{ Output}: From the FC layer,  $y^2_k=\sum_{n=0}^{5} o^1_F(n)w^2_k(n)$. \\
			$\mathbf{y}^2= (\mathbf{w}^2)^T\mathbf{o}^1_F=\begin{bmatrix} -0.38     &    -0.77
			\end{bmatrix}^T$.  \\
			\hline
			$\bullet$\textbf{ Softmax}: With $S=2$,  \  $P_k=e^{y^2_k}/(e^{y^2_1}+e^{y^2_2})$, \ $k=1,2$, we get \\
			$\mathbf{P}=\begin{bmatrix}
			0.60 & 0.40 \end{bmatrix}^T$,  \ \ \ 
			$\boldsymbol{\Delta}^2=\mathbf{P}-\mathbf{t}=\begin{bmatrix}
			-0.40 & 0.40 \end{bmatrix}^T$, \ \ $\Delta^2_k=P_k-t_k$ \\
			\hline
		\end{tabular}
		
	\end{centering}

	\begin{tabular}{|l|l|}
		
		\hline 
		\textbf{Back-propagation}: Delta error, gradient, weight update \\
		\hline
		\\
		$\bullet$\textbf{ Gradient} in the FC layer, $g^2_k(m)=\Delta^2_ko^1_F(m)$, \\
		$\mathbf{g}^2=\mathbf{o}^1_F(\mathbf{\Delta}^2)^T=\begin{bmatrix}
		-0.12 &-0.12 &-0.00 &-0.09 &-0.12 &-0.17 \\
		\phantom{-}0.12 &\phantom{-}0.12 &\phantom{-}0.00 &\phantom{-}0.09 &\phantom{-}0.12 &\phantom{-}0.17 
		\end{bmatrix}^T$,  \\
		\hline
		$\bullet$\textbf{ Weight update} in the FC layer,
		$w_k^2(m)\leftarrow w_k^2(m)+0.1g^2_k(m)$.\\
		$\mathbf{w}^2=\mathbf{w}^2-0.1\mathbf{g}^2=\begin{bmatrix}
		-0.45 &-0.04 &-0.05 &-0.79 &\phantom{-}0.64 &-0.15 \\
		\phantom{-}0.00 &\phantom{-}0.10 &-0.15 &-0.08 &-0.89 &-0.56 
		\end{bmatrix}^T$, \\
		\hline
		$\bullet$\textbf{ Delta error back-propagation} in the convolutional layer,  $\Delta^1_k(m)=\sum_p\Delta^2_pw^2_p(m)$, \\
		$(\boldsymbol{\Delta}^2)^T\mathbf{w}^2=[0.18, \ 0.06, \ \textbf{-0.04}, \ 0.29, \ -0.62, \ -0.16]$,  \\
		Repositioned \{$(\boldsymbol{\Delta}^2)^T\mathbf{w}^2$\}$=\begin{bmatrix}\phantom{-}0.18 & \phantom{-}0.06 \\ -0.04 & \phantom{-}0.29 \\ -0.62 & -016\end{bmatrix}$, following $\mathbf{o}_F^1 \rightarrow \mathbf{o}^1$ \\
		\hline
		$\bullet$\textbf{ Repositioning} the elements of $\boldsymbol{\Delta}_k^1=[(\boldsymbol{\Delta}^2)^T\mathbf{w}^2 \text{ at the positions defined by }\mathbf{M}^{MP}_k \odot\mathbf{M}^{ReLU}_k$  \\
		$\boldsymbol{\Delta}_1^1=\begin{bmatrix}0 & 0.18 & 0& 0&  0& 0.06\end{bmatrix}^T$,    \\
		$\boldsymbol{\Delta}_2^1=\begin{bmatrix}\textbf{0}& 0 \ \ \ \ \ \ & 0& 0& 0& 0.29\end{bmatrix}^T$  \\
		$\boldsymbol{\Delta}_3^1=\begin{bmatrix}0& \!\!\!\!\!\!\!\! -0.62& 0& 0& 0& \!\!\!\!\!\! -0.16\end{bmatrix}^T$ \\
		where $\odot$ is the Hadamard element-by-element product.\\ 
		\hline
		$\bullet$\textbf{ Gradient} in the convolutional layer, $g^1_k(m)=\Delta^1_k(m)*_cx(m)$, \ \ $\mathbf{g}^1_k=\boldsymbol{\Delta}_k^1 *_c \mathbf{x}$ \\
		$\mathbf{g}^1_1=\begin{bmatrix}-0.03&        \phantom{-}  0.03& \      \phantom{-}   0.06\end{bmatrix}^T$, \\ 
		$\mathbf{g}^1_2=\begin{bmatrix}\phantom{-}0.01&         \phantom{-}0.20& \        \phantom{-}  0.12\end{bmatrix}^T$,   \\
		$\mathbf{g}^1_3=\begin{bmatrix}\phantom{-}0.06&    -0.02&      \phantom{-}   0.06\end{bmatrix}^T$ \\
		\hline
		$\bullet$\textbf{ Weight update} in the convolutional layer $w_k^1(m)\leftarrow w_k^1(m)-0.1g_k(m)$ \\
		$\mathbf{w}^1_1=\begin{bmatrix}
		-0.06 & -0.01 & -1.47 \end{bmatrix}^T$ \\
		$\mathbf{w}^2_1=\begin{bmatrix}
		\phantom{-}0.45 & \phantom{-}0.13 & -0.31 \end{bmatrix}^T$, \\
		$\mathbf{w}^3_1=\begin{bmatrix}
		\phantom{-}1.12 & -1.01 & -1.84 \end{bmatrix}^T$  \\
		\hline
		$\bullet$\textbf{ Bias update}, $b_k \leftarrow b_k-0.05\sum_n \Delta_k^1(n)$ \\
		$\mathbf{b}\leftarrow \mathbf{b}-0.05(\begin{bmatrix}1&1&1&1&1&1\end{bmatrix} \boldsymbol{\Delta}^1)^T = \mathbf{0}-0.05\begin{bmatrix}
		0.24 & 0.29 & -0.78  \end{bmatrix}^T$. \\
		\hline
		$\bullet$\textbf{ New iteration} with the new signal,  $\mathbf{t}=[0, \ \ 1]$, \\
		$\mathbf{x}=\begin{bmatrix}
		-0.10 & -0.12 & -0.05 & -0.24 & 0.89 & -0.35 & -0.02 & -0.00 \end{bmatrix}^T$, \\
		Go back to the first step with the new (updated) weights, $\mathbf{w}^1$ and $\mathbf{w}^2$, and bias $\mathbf{b}$. \\
		\hline
	\end{tabular}
	
	\bigskip 
	
	\noindent(a) After the first training cycle is finished, as outlined step-by-step in the table above, the process is repeated with a new input noisy signal $\mathbf{x}$, randomly assuming $feature_1$ or $feature_2$, at a random position within the signal, as shown in Fig. \ref{Example_CNN}(a). 
	
	The following parameters during the training process are given in Fig. \ref{Example_CNN}:
	\begin{itemize}
		\item 
	The obtained probabilities, $P_k$, $k=1,2$, at the output of the CNN (being the output of the Softmax layer) are given in the second panel of Fig. \ref{Example_CNN} using black "+" for the values when the correct result should be $P_k=1$, that is, the value of $P_1$ is shown when the $feature_1$ is present and the value $P_2$ is given by this mark when the $feature_2$ is present in the input signal. In an ideal case all black "+" should be in positions where this value is equal to 1. The output values $P_k$ are designated by green ".", when the correct output result should be $P_k=0$, that is, this mark is used for $P_1$ when the $feature_2$ is present and for $P_2$ when the $feature_1$ is present in the input signal. In an ideal case all green "."  should be in positions where this value is equal to 0. The output signals (probabilities) are presented using marks "." and "+" in such a way that the correct positions of the mark "." would always be 0 and the correct position of the mark "+" would always be 1. 
	   
	   \item The values of weights in the fully connected layer, during the training process, are given in the third panel in Fig. \ref{Example_CNN}.  
	   
	   \item The training process is performed using $200$ random realizations of the input signal, randomly assuming $feature_1$ or $feature_2$. This cycle of $200$ realizations is called an epoch. Then the same set of $200$ random realizations is repeated  $10$ times (then epochs are used in training), that is the CNN  was trained over 10 epochs, with no max-pooling  used. 
	   
	   \item After the CNN is trained in 10 epochs of 200 random realizations of the signal, the update process of the weights in all layers is stopped, and the achieved weights are tested on 100 new random realizations. The results are shown in the last panel in Fig. \ref{Example_CNN}. We can see that the decision was correct in all 100 new cases, where the marks black  "+" and green "." are used in the sense described in the first item of this list.    	      
	   \end{itemize}
	
	\noindent(b)  The same setup  in  Fig. \ref{Example_CNN} was next used in a CNN with the max-pooling operation, using the factor $P=3$. The results are shown in  Fig. \ref{Example_CNN_P3} with the same explanation as in Fig. \ref{Example_CNN}. 	Observe that without max-pooling, the probabilities separate after 600 iterations (3 epochs), while in the case with max-pooling 800 iterations (4 epochs) are needed. 
	
	The number of weights when no max-pooling is used, was $((N-M+1)K) \times 2 = 24\times2=48$, while with max-pooling it was $((N-M+1)K/P) \times 2 = 6\times2=12$. In the case without max-pooling, we used $K=4$ channels, while in the case of max-pooling, the number of channels was reduced to $K=3$.
	
	\noindent(c)  Finally, the same signal was used two train a CNN with one convolutional layer and two fully connected layers, with $K=5$ channels in the convolution and $P=3$ being used in max-pooling operation. The number of input neurons in the second fully connected layer was $N_2=4$. The Softmax with two output neurons was used for the decision. The results are shown in Fig. \ref{Example_CNN_2_P3_3layers}, with the same notation as in the previous figures. We can see that the convergence of the weights was faster than in the previous two cases. After 300 training cycles the weights assumed almost steady values. The testing of this network on 100 new random realizations of the input signal was 100\% successful, as observed from the last panel in Fig. \ref{Example_CNN_2_P3_3layers}.  
	
\end{exmpl}     

\begin{figure}[hptb]
	\begin{center}
		\includegraphics{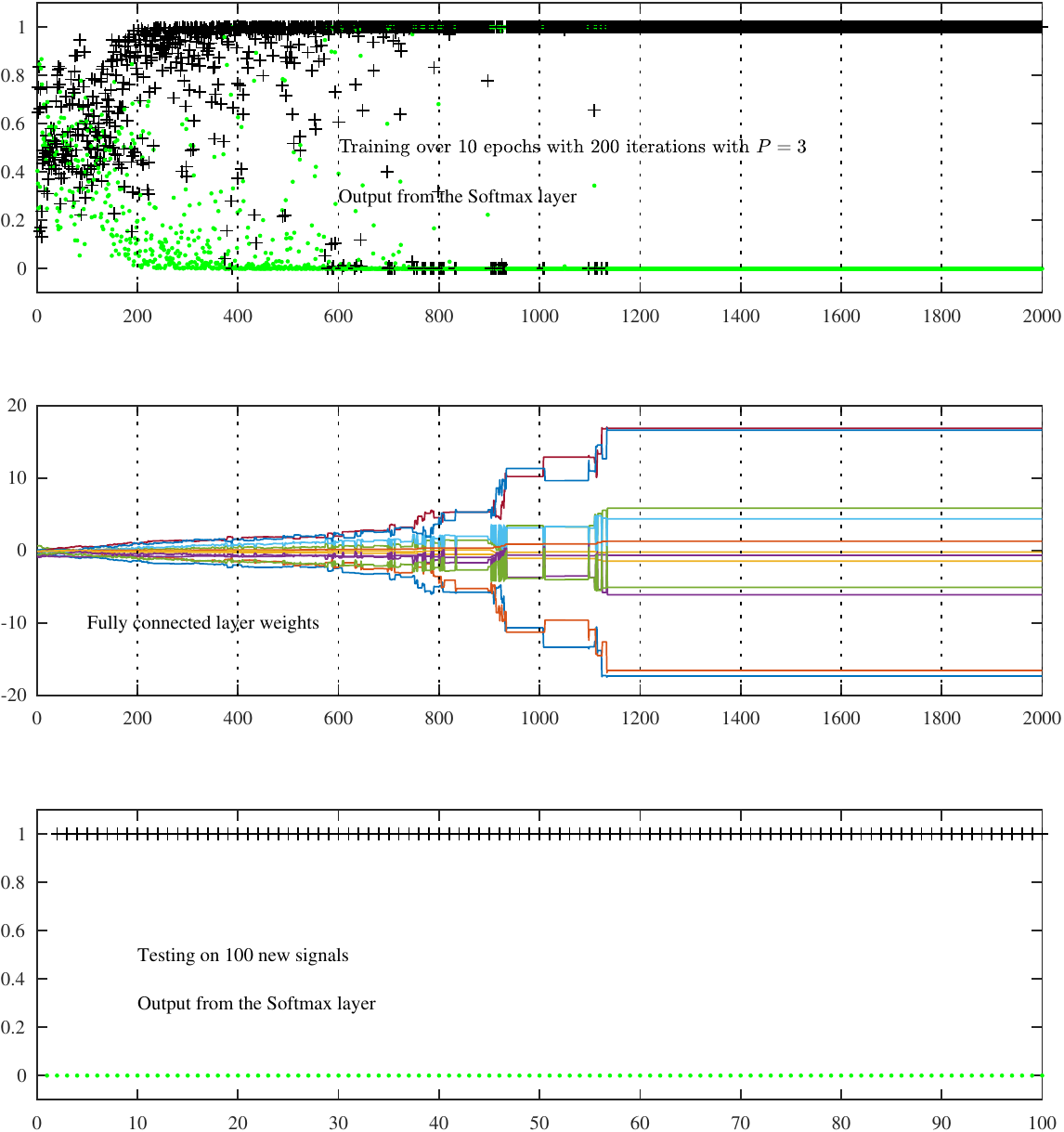} 
		\caption{Illustration of the operation of a a CNN with one convolutional layer and one FC layer, with two neurons at the output (Softmax) layer, max-pooling with $P=3$, and $K=3$ channels. This helped reduce the number of weights in the FC layer to only $((N-M+1)K/P) \times 2 = 6\times2=12$ weights.}%
		\label{Example_CNN_P3}%
	\end{center}
\end{figure}

\begin{figure}[hptb]
	\begin{center}
		\includegraphics{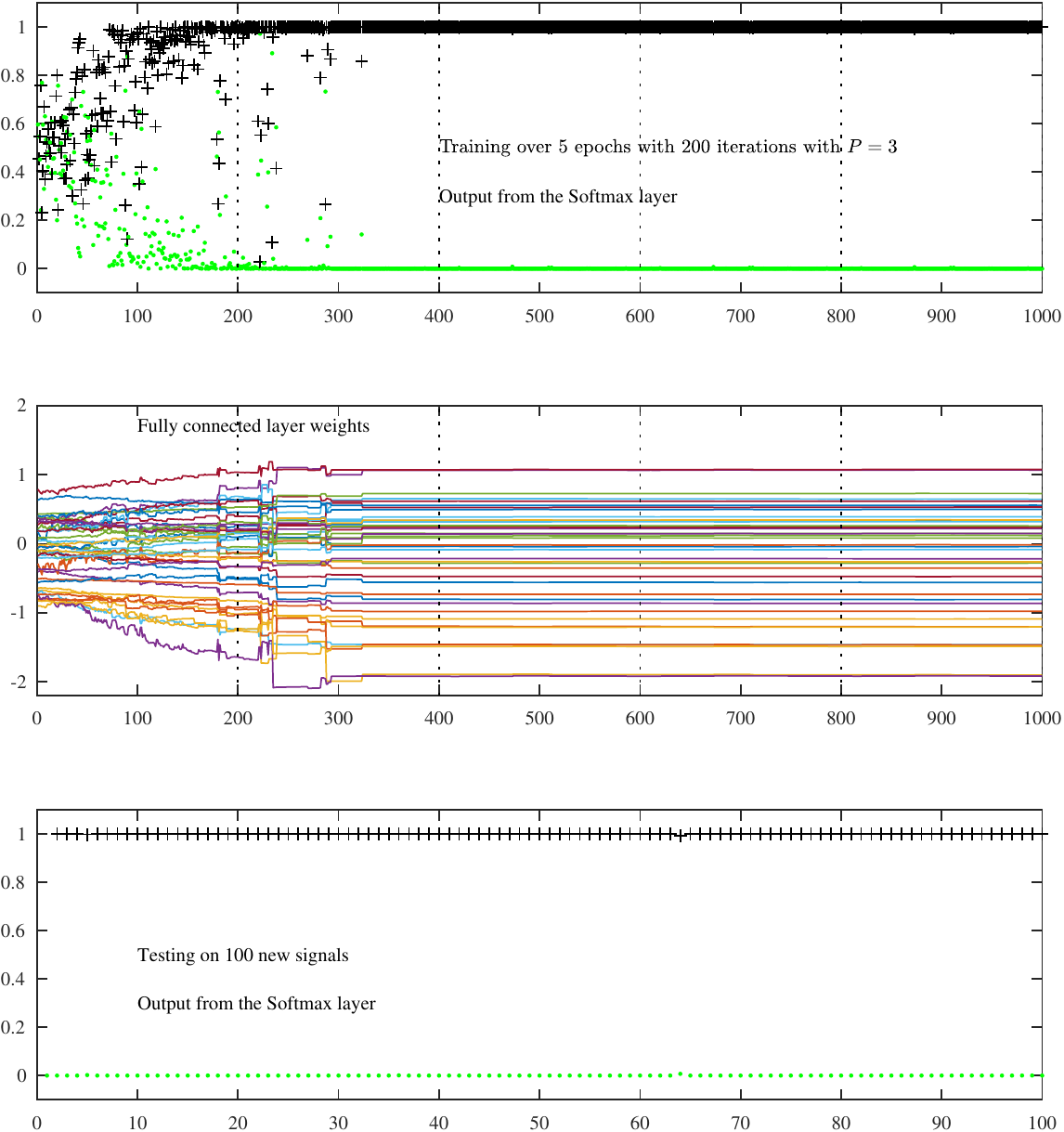} 
		\caption{Illustration of the operation of a CNN from Fig. \ref{fully_connected_layer11}  with one convolution layer and \textit{two} FC layers, with two neurons at the output (softmax) layer, max-pooling with $P=3$, and $K=5$ channels. The number of  neurons between two FC layers is $N_2=4$. This reduces the number of weights in the FC layer to $((N-M+1)K/P) \times N_2+N_2\times 2 =48$ weights. High training accuracy was achieved with 5 epochs over 1000 iterations. The evolution of weight updates in the first FC layer are shown in the middle panel, while the bottom panel shows the overall network output for test data.}%
		\label{Example_CNN_2_P3_3layers}%
	\end{center}
\end{figure}

\newpage

\section{Additional Considerations}

\noindent \textbf{Convolutions with $1$ in signals or $1 \times 1$ filters in images.}  From Fig. \ref{CNN_conv2} we can see that the number of weights in filters is increased $K$ times for $K$  output signals $o^l_p(n)$, for channels $p=1,2,\dots,K$. This new channel dimension increases the number of weights $K$ times. Therefore, the number of parameters increases linearly with the number of convolutions (filter patterns), $K$, but can be reduced using the so called $1$ filters in signals or $1 \times 1$ filters in images. We will consider here the simplest and the most commonly used case that reduces this dimension from $K$ to 1. To this end consider Fig. \ref{CNN_conv2}, with the filter of length $M=1$ and a given $k$, with the weights $w^2_{k,p}(0)$, $p=1,2,\dots,K=4$, for every $k=1,2,\dots,K_2$. Then, we obtain just one dimensional output $y_k^2(n)$, $k=1,2,\dots,K_2$. Next, $K_2$ filters be applied, as in Fig. \ref{CNN_conv}, to this one-dimensional signal, to produce the resulting convolutional output. This approach is called \textit{convolutions with $1$} and  may significantly reduce the number of required weights. Indeed, the  number of weights for filters of width $M_2$ was $M_2KK_2$, while if  convolutions with $M_2=1$ are used first, then initially we have $KK_2$ filters 1, to reduce the dimension $K$ to 1, and then $M_2K_2$ filters for the convolution of signals obtained in such a way. In total, the reduction is significant, since $KK_2+M_2K_2=(M_2+K)K_2 < (M_2K)K_2$.

Of course, we may use different lengths of the $1 \times 1$ filter in the direction $p$ to reduce, or even increase, the number of weights (if zero-padding is added). 

\bigskip

\noindent \textbf{Dropout for Regularizing Deep Neural Networks.}
Given a large number of neurons and layers, deep neural networks are likely to quickly \textit{overfit} a training dataset. Within the help of the convolutional layers of a CNN, this problem is reduced through max-pooling or output down-sampling (stride). In both cases, the outputs which are ignored in the next layer are either defined by the signal and filters (max-pooling) or by a regular down-sampling scheme (stride). In deep neural networks, regularization is commonly achieved by \textit{randomly dropping out nodes} in the network, whereby every node is considered as a candidate to be dropped out (ignored) with probability $P$. 
Then, a deep neural network is trained by means a large number of neural networks, with different architectures, operating in parallel, obtained by different random dropped nodes.

The effect of the neuron drop-out is such that during the training,  each update within a layer is performed with a different "view" of the configured layer, forcing nodes within a layer to probabilistically take on more or less responsibility for the inputs and co-adapt to correct mistakes from the previous layers, in turn making the model more robust. In this sense, neuron dropout represents a kind of a sparse activation function from a given layer. 

A CNN with dropout can be implemented in the same way as the above described approaches. Notice that since several networks are trained in parallel, a normalization of the weights from each architecture, with probability $P$, should be performed.

\section{Conclusion}
We have employed the matched filtering paradigm as a ``mathematical lens'' to demystify the operation and learning in Convolutional Neural Networks (CNN). A close examination of the convolutional layer within CNNs has revealed a direct and intuitive link with matched filtering for finding the features (patterns) in data. Such a framework has allowed us for a seamless transition between matched filtering and feature identification, together with a unifying and a straightforward platform for understanding the information flow in learning and optimal parameters selection.  The material is supported by a comprehensively evaluated example, with detailed numerical outputs and visualizations. This approach has been shown to permit the introduction of CNNs in a theoretically well founded and physically meaningful way, which is beneficial for many communities that do not rely on black box approaches. In addition, the material may be useful in lecture courses in statistical signal processing, machine learning, and statistics, or indeed, as an interesting step-by-step guide to CNNs for the intellectually curious and generally knowledgeable reader.

\newpage

%\nocite{*}
\bibliographystyle{ieeetr}

\bibliography{graph-signal-processing}

\end{document}